\documentclass[twocolumn,prb,nopacs,amsmath,amssymb]{revtex4-1}
\usepackage[dvips]{graphicx}
\usepackage{dcolumn}% Align table columns on decimal point
\usepackage{bm}% bold math
\usepackage{color}
\def\be{\begin{equation}}
\def\en{\end{equation}}
\def\bea{\begin{eqnarray}}
\def\ena{\end{eqnarray}}
\def\bec{\begin{equation}\begin{array}{rcl}}

\def\p{\partial}
\def\ep{\epsilon}
\def\gs{\gtrsim}
\def\ls{\lesssim}

\def\ve{\varepsilon}
\newcommand{\av}[1]{\langle{#1}\rangle}

\newcommand{\bi}[1]{\mbox{\boldmath$#1$}}

\newcommand{\ten}[1]{\stackrel{\leftrightarrow}{\bi{#1}}}

\def\a1{\stackrel{\leftrightarrow}{1}}

\begin{document}
\title{Ferroelectric glass of 
 spheroidal dipoles with   impurities: 
Polar nanoregions,\\ response to applied electric field, 
and ergodicity breakdown }  
\author{Kyohei Takae$^1$ and Akira Onuki$^2$ }
\affiliation{
$^1$Institute of Industrial Science, University of Tokyo,
4-6-1 Komaba, Meguro-ku, Tokyo 153-8505, Japan\\
$^2$Department of Physics, Kyoto University, Kyoto 606-8502, Japan}
%\email{onuki@scphys.kyoto-u.ac.jp}
\date{\today}

\begin{abstract} 
Using molecular dynamics simulation, we study dipolar glass 
in crystals composed of slightly spheroidal,  polar 
particles and spherical,  apolar impurities between  metal walls. 
We present physical pictures   of ferroelectric glass, 
which have been observed in   relaxors,   mixed crystals (such as 
KCN$_x$KBr$_{1-x}$), and polymers. Our systems undergo  a  diffuse 
 transition in a wide temperature range, 
where we visualize  polar nanoregions (PNRs)  surrounded by impurities. 
In our simulation, the impurities  form clusters and 
their space distribution is  heterogeneous. 
The polarization fluctuations are   enhanced  at relatively high   
$T$ depending on the size of the dipole moment. 
They then form  frozen PNRs  
as   $T$  is further lowered into  the  nonergodic regime. 
As a result, the dielectric permittivity  exhibits  
 the  characteristic features of  relaxor ferroelectrics. We also examine  
nonlinear response to cyclic applied electric field 
and  nonergodic response to cyclic temperature changes (ZFC$/$FC), where 
the polarization and the strain change collectively  and 
heterogeneously. We also study antiferroelectric glass 
arising from molecular shape asymmetry. 
We use  an Ewald  scheme of calculating the dipolar interaction 
in applied electric field.  
\end{abstract}

%\pacs{64.70.K-, 61.72.-y, 77.80.Jk, 77.65.-j}
%\keywords{}

\maketitle

\section{Introduction}

Ferroelectric transitions  have been  
attracting much attention in various systems. 
It is known that they can occur even in simple particle systems.  
For example,   one-component spherical particles with a point dipole 
 undergo a ferroelectric  transition 
 in crystal or  liquid-crystal phases if the dipole interaction 
is sufficiently strong  
\cite{Gao,Wei,Weis,Ta,Tao,Ayton,Hent0,Hent,Dij0,Groh}. 
Such spherical dipoles  form various noncubic crystals in  ferroelectric phases \cite{Groh,Dij0}.  Ferroelectriciity was also  studied in positionally 
disordered dipolar solids\cite{Ayton}. 
Recently, Johnson {\it et al.}\cite{Johnson0,Johnson1} 
have investigated  a  ferroelectric transition 
of  spheroidal particles with a dipole moment $\mu_0$ 
parallel to the spheroidal axis.
 They found that  the static dielectric constant increases 
up to  $10^2-10^3$ with increasing $\mu_0$ 
if the aspect ratio is  close to unity.  
In this paper, we examine   ferroelectric transitions 
in mixtures of slightly  spheroidal dipoles and spherical impurities.  
%where disorder is crucial.

In many solids,  the polarization 
is   induced by ion  displacements within unit cells 
and the dielectric constant  is  very large. 
As a unique aspect, the ferroelectric 
transitions become {\it diffuse }
  with a sufficient amount of disorder\cite{Vug,Bl,Kob}, 
  which take place   over a wide temperature range 
without long-range dipolar order. Notable examples  are relaxors 
\cite{Smo,Klee,Klee1,Bokov3,Bokov,Bl,Cr,Cowley,Samara} 
such as Pb(Mg$_{1/3}$Nb$_{2/3}$)O$_3$ (PMN), 
where   the random  distribution of  Mg$^{2+}$ and Nb$^{5+}$ at B sites 
yields quenched random fields at Pb$^{2+}$  sites 
\cite{Setter,West}. 
In relaxors, temperature-dependence of the optic index 
of refraction  suggested appearance  of mesoscopic 
polarization  heterogeneities \cite{Burns}, 
called  polar nanoregions (PNRs). 
They are enhanced  at relatively high $T$ 
as near-critical fluctuations 
and are frozen at lower $T$. 
It is widely believed that  these  PNRs give rise to a broad peak   in the 
  dielectric permittivity $\ve'$ as a function of $T$
\cite{Setter,Smo,Klee,Klee1,Bl,Cr,Cr1,Bokov,Bokov3,Cowley,Samara,Ra}. 
 They  have been detected  by neutron and x-ray 
scattering \cite{Wel,Stock,Eg,Shi1,Shi2,Cowley} 
and visualized by  transmission electron microscopy \cite{Ban,Mori,Uesu} 
and piezoresponse force microscopy \cite{Bokov2,Bokov4,Ann}. 
Strong correlations have also been found between the PNRs and 
  the compositional  heterogeneity  of the B site cations
\cite{Wel,Wel1,Ban,Setter,Bokov4,Mori,Perrin,Jin,i3,i6}.

Relaxor  behaviors  also appear 
in other disordered dipolar  systems\cite{Bl,Kob,Vug}. 
In particular, orientational  glass has 
long been studied  in mixed crystals 
such as KCN$_x$KBr$_{1-x}$  or K$_x$Li$_{1-x}$TaO$_3$ 
\cite{Vug,Tou,Ho1,Ho2,Loidl3,qua,ori,Binder,anti-Loidl,Yokota,Mertz,F3}, 
where the two mixed  components   have 
similar sizes and  shapes. Upon    cooling  below   melting, 
they first  form a cubic crystal without long-range orientational  order 
in the {\it plastic crystal}  phase. 
At lower  $T$, an  orientational phase transition  
 occurs,  where the crystal structure becomes 
noncubic. In  nondilute 
mixtures, this  transition  is  diffuse 
with slow relaxations, 
where   the orientations and the strains are strongly coupled, both 
exhibiting   nanoscale 
heterogeneities\cite{qua,Heuer,Takae-ori,EPL}. 
Some polymers   also undergo 
 ferroelectric transitions 
 due to alignment of permanent dipoles \cite{Bl,Lov,Furukawa,Furu1}. 
In particular, 
poly(vinylidene fluoride-trifluoroethylene) copolymers\cite{Zhang,Z1}  
exhibited large electrostriction 
and    relaxor-like  polarization responses 
after electron irradiation 
(which brings disorder in polymer crystals).  
We also mention strain glass  in  shape-memory   
 alloys \cite{F5}, where  the    dipolar interaction 
does not come into play but a  diffuse ferroelastic 
transition  occurs with strain heterogeneities.  
 We   now recognize the universal  
 features of  glass coupled with  a  phase transition, 
where   the order parameter fluctuations  are frozen 
at low $T$.

In  their molecular dynamics 
 simulation  of  relaxors,  Burton 
{\it et al}.\cite{Burton1,Burton2,Burton3}  
  started with  a first-principles  Hamiltonian 
for  atomic  displacements  in perovskite-type  crystals. 
As a compositional distribution,   
 they assumed nanoscale  chemically 
ordered regions embedded in a chemically  
disordered matrix.   On the other hand, we   investigate general 
aspects   of  ferroelectric glass with 
a simple molecular model.  In electrostatics, we use   an Ewald scheme 
 including  image dipoles and applied 
electric field \cite{apply,Klapp}, 
 which has been  used  to study 
 water between  electrodes\cite{Hautman,Takae1,Takae2}.
To prepare  a mixed crystal, we 
 cool   a liquid mixture from high $T$; then, 
our  impurity distribution at low $T$ 
 is naturally formed   during crystallization 
\cite{Takae-ori,EPL}.

Our system  consists  of  spheroidal dipoles   and 
 spherical apolar  particles only. Nevertheless, 
  we can realize enhanced  polarization 
fluctuations forming   PNRs and 
 calculate the frequency-dependent 
dielectric permittivity. We can also calculate the  
 responses to   applied electric field 
and  to  ZFC$/$FC (zero-field-cooling and field-cooling)  
temperature changes. In the latter,  nonergodicity 
of glass is demonstrated, so its experiments  have  been 
performed  in spin glass\cite{Mydosh,F0,F1}, relaxors\cite{West,Uesu,F4}, 
orientational glass\cite{Ho1,F3}, relaxor-like polymers \cite{Z1}
, and strain glass\cite{F5}.

The organization of this paper is as follows. In Sec. II,
we will explain our theoretical scheme 
and numerical method.  In Sec. III, we will 
explain a structural phase transition in 
a one-component system of dipolar spheroids. 
In Sec. IV, we will examine 
diffuse ferroelectric transitions  with impurities. 
Furthermore, we will examine responses to  cyclic applied field in Sec.V 
and to  cyclic temperature changes in Sec.VI. 
Additionally,  antiferroelectric 
glass will be briefly discussed In Sec.VII.

\section{Theoretical background}

We treat  mixed crystals composed of 
spheroidal polar  particles as the first species and 
spherical apolar particles (called impurities) as 
the second species. 
These particles have no electric charge. 
As in Fig.1(a), we suppose smooth metal walls at $z=0$ and $H$ 
to apply electric field to the dipoles.  The periodic boundary  
condition is imposed  along the $x$ and $y$ axes with period $L$. 
Thus,  the particles are in 
  a $L\times L\times H$ cell with  volume  $V=L^2H$.

In terms of the impurity concentration $c$, 
the particle  numbers of the two species  are written as 
\be  
N_1=Vn_1= (1-c)N,\quad N_2=Vn_2=cN, 
\en 
where the total particle number $N$ 
 is set equal to $8000$. Their  positions are 
written as ${\bi r}_i=(x_i,y_i,z_i)$ ($1\le i\le N$). 
The  long  axes of the spheroidal particles 
are  denoted by unit vectors 
 ${\bi n}_i=(n_{xi},n_{yi},n_{zi})$ 
($1\le i\le N_1$).

\begin{figure}[tbp]
\begin{center}
\includegraphics[width=200pt]{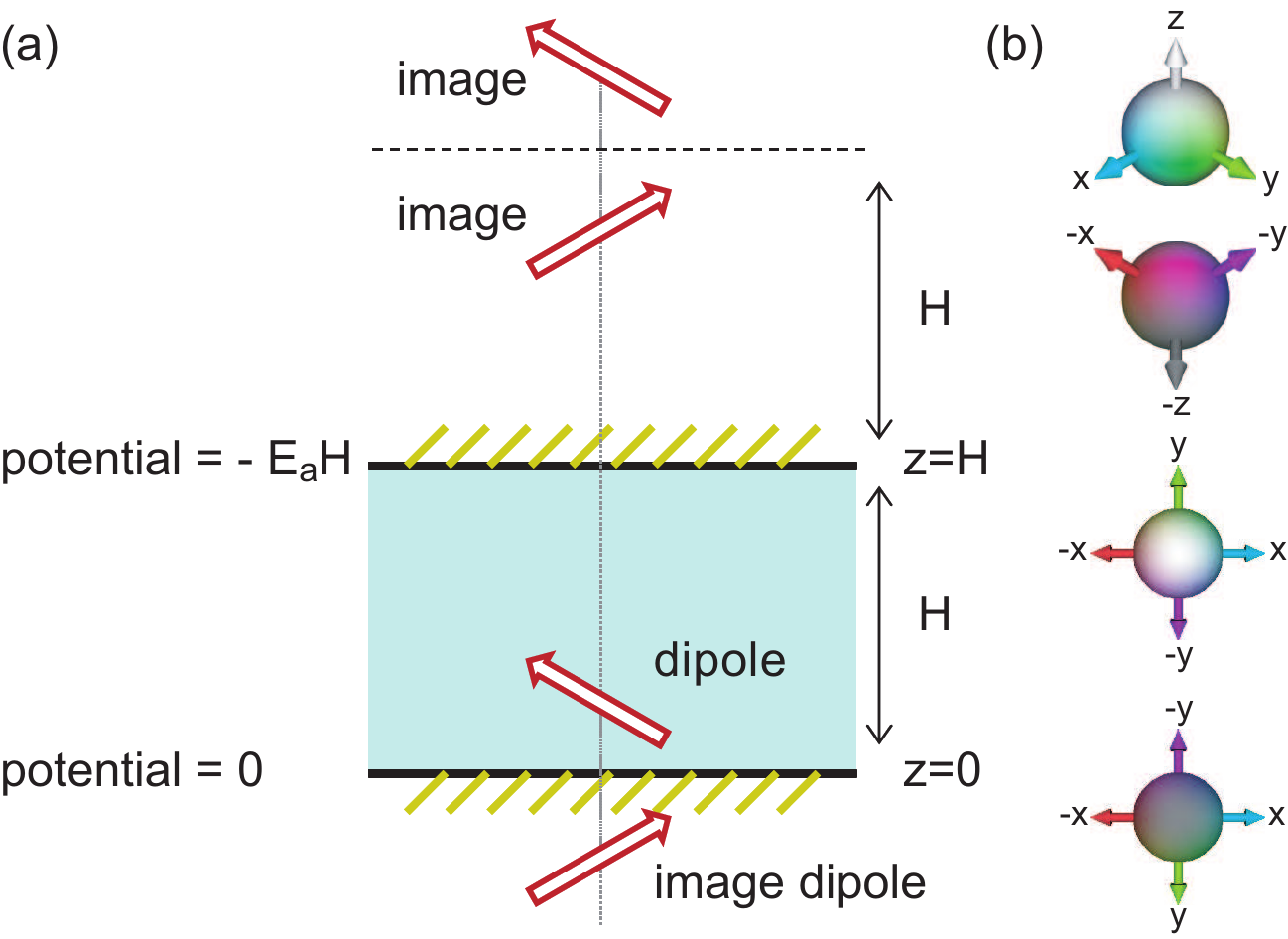}
\caption{  
(a) Illustration of geometry. 
Dipoles are in a cell (green region) 
and two kinds of image dipoles are outside it. 
Parallel metallic plates are at $z=0$ and $H$. 
Electric  potential  is 0 at $z=0$ and is 
$-\Delta\Phi= -E_{\rm a} H$  at $z=H$. 
(b) Color maps of dipole orientation  
on a sphere surface used in the 
following figures. 
Displayed  from above are diagonally downward view, 
diagonally upward view, 
top view, and bottom view. 
}
\end{center}
\end{figure}

\subsection{Potential energy}

The  total potential energy $U$ is expressed as 
\be
U =U_{\rm LJ}  +U_{\rm w} +U_{\rm d}. 
\en 
Here, $U_{\rm LJ} $ is the sum of modified 
Lennard-Jones potentials between particles 
$i\in\alpha$ and $j\in\beta$ ($\alpha,\beta =1,2$),  
\be 
U_{\rm LJ}  =2\ep\sum_{i\neq j} \bigg[
(1+A_{ij})\frac{\sigma_{\alpha\beta}^{12}}{r_{ij}^{12}}-
\frac{\sigma_{\alpha\beta}^6}{r_{ij}^6}\bigg]. 
\en
 where $r_{ij}= |{\bi r}_i-{\bi r}_j|$, $\epsilon$ is the characteristic 
interparticle energy, 
and  $\sigma_{\alpha\beta}=(\sigma_\alpha+\sigma_\beta)/2$ 
in terms of the   particle lengths  
$\sigma_1$ and $\sigma_2$. 
The factor $A_{ij}$  depends      on the angles between spheroid 
 directions  and ${\bi r}_{ij}= {\bi r}_{i}-{\bi r}_j$ as
 \cite{EPL,Takae-ori}
\be
A_{ij}= \delta_{\alpha 1} {\eta} ({\bi n}_i\cdot{\bi r}_{ij}/r_{ij})^2
+\delta_{\beta 1} {\eta} ({\bi n}_j\cdot{\bi r}_{ij}/r_{ij})^2,   
\en 
where $i\in 1$ in the first term, $j\in 1$ in the second term, 
and  $\eta$ represents the molecular anisotropy. 
For $\eta>0$, we have $0\le A_{ij}\le 2\eta$, which  vanishes for 
${\bi n}_i\cdot{\bi r}_{ij}=0$ for $j\in 2$ 
(and ${\bi n}_j\cdot{\bi r}_{ij}=0$ for $j\in 1$). 
We assume   a relatively  small size difference and mild anisotropy as 
\be 
\sigma_2/\sigma_1=1.1, \quad \eta =1.2. 
\en 
Then,   at density $0.84\sigma_1^{-3}$, 
our system  forms a crystal without phase separation 
and isotropic-nematic phase transition \cite{Latz,EPL}. 
For larger $\sigma_2/\sigma_1$ and $\eta$, the latter  processes  
may take place  during slow quenching from liquid.  
Because $U_{\rm LJ}$ is minimized at $r_{ij}= 
(2+ 2A_{ij})^{1/6}\sigma_{\alpha\beta}$ 
for fixed  ${\bi n}_{i}$ and ${\bi n}_j$, 
  we regard the anisotropic particles as spheroids with 
aspect ratio $(1+ 2\eta)^{1/6}=1.24$. 
Notice  that our   potential is similar
to  the Gay-Berne potential  
for rodlike molecules \cite{Berne}.

The second term  $U_{\rm w}$ in Eq.(2) 
is the sum of  strongly repulsive, wall potentials as \cite{apply}    
\be 
U_{\rm w}=w \sum_i[ \exp(-z_i/\xi_{\rm w})+\exp(-(H-z_i)/\xi_{\rm w})],
\en 
We set   $w=e^{40}\epsilon$ and  $\xi_{\rm w} =0.01\sigma_1$ to make 
the potentials hardcore-like. 
Then,  the distances between 
the dipole centers and  the walls become 
longer than   $0.4\sigma_1$.

\subsection{Electrostatic energy and canonical distribution} 

We assume   permanent dipolar moments ${\bi \mu}_i$  
along  the spheroid direction 
${\bi n}_i$ ($1\le i\le N_1$) written as 
\be 
{\bi\mu}_i=(\mu_{xi},\mu_{yi}, \mu_{zi})=\mu_0{\bi n}_i,
\en  
where  $\mu_0$ is a constant dipole moment. There is 
no induced dipole moment. The electric potential 
$\Phi({\bi r})$ can be  defined away from the dipole positions 
${\bi r}\neq {\bi r}_i$. We impose the 
metallic boundary condition at $z=0$ and $H$:  
\be 
\Phi(x,y,0)=0 , \quad \Phi(x,y,H)= -\Delta\Phi= -E_{\rm a} H, 
\en 
where   $\Delta\Phi$  is   the applied potential difference 
and  $E_{\rm a}=\Delta\Phi/H$ is  the applied electric field. 
In this paper, we perform simulation by controlling 
$\Delta\Phi$ (or $E_{\rm a}$). In our scheme, $\Delta\Phi$ 
can be nonstationary.

The boundary condition (8) is   realized 
by  the surface charge densities at $z=0$ and $H$ (see Appendix A). 
As a mathematical convenience,  we instead 
 introduce  image dipoles  outside the cell 
 for   each dipole  ${\bi\mu}_i$ at ${\bi r}_i=(x_j,y_j,z_j)$ in the cell. 
As in Fig.1(a),    we first consider 
those  at ${\bi r}_i-2H m_z{\bi e}_z$   ($m_z=\pm 1,\pm 2, \cdots$) 
   with the same moment  ${\bi\mu}_i$, 
where ${\bi e}_z$ is the unit vector along the $z$ axis. 
Second,    at 
 ${\bar{\bi r}}_i- 2H m_z {\bi e}_z$  $(m_z=0, \pm 1, \pm 2,\cdots) $, 
we consider  those with the image   moment given by 
\be 
\bar{\bi\mu}_i= 
(-\mu_{xi}, -\mu_{yi},\mu_{zi}),
\en 
where ${\bar{\bi r}}_i= (x_i,y_i, -z_i)$ is the image position closest 
to the bottom wall. For ${\bi r}\neq {\bi r}_i$, 
 the real and image dipoles and the applied field yield the following potential, \bea 
\Phi({\bi r}) &=&  \sum_{{\bi h}} {\sum_{j\in 1}} \bigg
 [ {\bi g} ({{\bi r}-{\bi r}_{j} +{\bi h}})\cdot{\bi\mu}_j  \nonumber\\
&& +  {\bi g} ({{\bi r}-{\bar{\bi r}}_{j}  
 + {\bi h}})\cdot{\bar{\bi \mu}}_j \bigg] -E_{\rm a}z ,
\ena 
where ${\bi r}\neq {\bi r}_i$, ${\bi g}({\bi r})= r^{-3}{\bi r}$,   and 
${\bi h}= (Lm_x, Lm_y, 2H m_z)
$ with $m_x, m_y$, and $m_z$ being integers. Here, 
the first term  is  periodic in 
three dimensions (3D). 
%along the $x$ and $y$ axes with  the  summation  over $m_x$ and $m_y$. 
Along the $z$ axis the   period is $2H$ 
 because of the   summation  over $m_z$ 
or over  the image dipoles. 
We confirm that the first term   in Eq.(10) vanishes   at $z=0$ and $H$ 
with the aid of  Eq.(9). 

At fixed $E_{\rm a}$, the   total electrostatic energy   $U_{\rm d}$ in Eq.(2) 
  is now written in terms of ${\bi r}_i$ 
and ${\bi \mu}_i$   as\cite{Klapp,apply}  
\bea 
&&
U_{\rm d}= \frac{1}{2} 
  \sum_{{\bi h}}{\sum_{i\in 1, j \in 1}}'
 {\bi \mu}_i \cdot {\ten{\cal T}}  
({\bi r}_{ij} +  {\bi h}) \cdot{\bi{\mu}}_j \nonumber\\
&&+  \frac{1}{2} 
  \sum_{{\bi h}}
 { \sum_{i\in 1, j \in 1}} {\bi \mu}_i \cdot {\ten{\cal T}}  
({\bar{\bi r}}_{ij} +  {\bi h}) \cdot{\bar{\bi{\mu}}}_j
 -E_{\rm a} M_z . 
\ena 
Here,  ${\ten{\cal T}}({\bi r})$ is the 
dipolar tensor with its $\alpha\beta$ 
component being $\delta_{\alpha \beta}/r^3-3 x_\alpha x_\beta/r^5$.
In the first term, 
%we set ${{\bi r}}_{ij}= {\bi r}_i-{{\bi r}}_j$  and 
the self-interaction contributions  $({\bi h}={\bi 0}$ and $i=j$) 
 are removed in $\sum_{i\in 1,j\in 1}'$. In the second term,   we set 
 ${\bar{\bi r}}_{ij}= {\bi r}_i-{\bar{\bi r}}_j=(x_i-x_j,y_i-y_j, 
z_i+z_j)$.  In the last term, $M_z$ 
is the $z$ component of the total polarization, 
\be 
{\bi M}= (M_x,M_y, M_z)= \sum_i{\bi \mu}_i. 
\en 
For each dipole $i$,  the electrostatic force  is given by 
${\bi F}_i^{\rm e}= - \p U/\p {\bi r}_i$ and 
the   local electric field  by 
\be 
{\bi E}_i= -\p U_{\rm d}/\p {\bi \mu}_i.
\en 
We can also obtain  
${\bi E}_i$  by subtracting the self 
contribution ${\bi g}({\bi r}-{\bi r}_i) \cdot {\bi \mu}_i$ 
from  $\Phi({\bi r})$  in Eq.(10)  as 
\be 
{\bi E}_i
=-\lim_{{\bi r}\to {\bi r}_i} 
\nabla [\Phi ({\bi r})-  {\bi g}({\bi r}-{\bi r}_i) \cdot {\bi \mu}_i ].
\en

We consider the  Hamiltonian ${\cal H}= {\cal K}+U$, where 
  ${\cal K}$ is   the total kinetic energy.  In our model,   the applied field $E_{\rm a}$ appears  linearly in $U_{\rm d}$ in Eq.(11). Then, we  find  
\be   
{\cal H}={\cal H}_0-M_zE_{\rm a}, 
\en 
where ${\cal H}_0$ is the Hamiltonian for $E_{\rm a}=0$. This  form was  
assumed in the original linear response theory\cite{Kubo}. 
For stationary $E_{\rm a}$, the equilibrium average, 
denoted by ${\av{\cdots}}_{\rm e}$, is  over the  canonical distribution 
  $\propto \exp(-{\cal H}/ k_{\rm B}T)$. 
Then, for any variable ${\cal A}$ 
(independent of $E_{\rm a}$),  its equilibrium average $\av{{\cal A}}_{\rm e}$ 
changes as a function of  $E_{\rm a}$ as\cite{Onukibook} 
\be 
\frac{\p}{\p E_{\rm a}} \av{{\cal A}}_{\rm e}= \frac{1}{k_{\rm B}T} \av{{\cal A}\delta M_z}_{\rm e},
\en 
where   $T$ is fixed in the derivative 
and $\delta M_z= M_z-\av{M_z}_{\rm e}$. For  the average 
polarization   $P_z=\av{M_z}_{\rm e}/V$, 
we consider   the differential susceptibility 
 $\chi_{\rm dif} = \p P_z/\p E_{\rm a}$. 
In equilibrium, it is related  
 to the variance of  $\delta M_z$  as
\be 
\chi_{\rm dif} = 
\frac{\p {P_z}}{\p E_{\rm a}}=\frac{1}{V k_{\rm B}T} 
\av{(\delta M_z)^2}_{\rm e}.
\en  
%which holds for any $E_{\rm a}$. 
 As $E_{\rm a}\to 0$,  $\chi_{\rm dif}$ tends to the  
susceptibility $\chi =(\ve-1)/4\pi$ in the linear regime. 
  In this paper, we calculate  the time averages 
 of the physical quantities  using data  from a single simulation run. 
In our case,  the ergodicity holds   at relatively high $T$,   but 
we do not obtain   Eq.(17) at low $T$ 
 because of  freezing of mesoscopic  PNRs  in our finite system 
(see Sec.IVC and  Fig.7).

\subsection{Kinetic energy and equation of motions}

The total kinetic energy $\cal K$ depends on 
the translational  velocities ${\dot{\bi r}}_i= {d{\bi r}_i}/{dt}$ 
($i=1,\cdots, N$)  and  the  angular velocities 
${\dot{\bi n}}_i= {d{\bi n}_i}/{dt}$ ($i=1,\cdots, N_1$) as 
\be 
{\cal K} = \frac{1}{2} \sum_{i} m |{\dot{\bi r}}_i|^2+ 
 \frac{1}{2}\sum_{ i\in 1} I_1 |{\dot{\bi n}}_i|^2,
\en 
where $m$ is the mass common to the two species,  
and $I_1$ is the moment of inertia.  
We set $I_1=0.125m\sigma_1^2$ in this paper. 
%We treat $U$    as 
%a function of ${\bi r}_i$ and ${\bi \mu}_i$.
 The Newton equations for ${\bi r}_i$ are given by 
\be 
m{\ddot{\bi r}}_i
 =- {\p U}/{\p{\bi r}_i },
\en 
where ${\ddot{\bi r}}_i={d^2{\bi r}_i}/{dt^2} $.  
On the other hand, the Newton equations for 
 ${\bi n}_i$   ($1\le i\le N_1$) are  of the form\cite{apply,EPL}, 
\be
 I_1({\ddot{\bi n}}_i + |{\dot{\bi n}}_i|^2 {\bi n}_i) 
= (\ten{I}- {\bi n}_i {\bi n}_i)\cdot \mu_0 {\bi E}_i^{\rm eff} ,  
\en 
where ${\ddot{\bi n}}_i= {d^2{\bi n}_i}/{dt^2}$,  
 $\ten{I}$ is the unit tensor, and  
${\bi E}_i^{\rm eff}= -\p U/\p {\bi \mu}_i$ is 
the local orientating field on dipole $i$. 
The left hand side of Eq.(20) 
is perpendicular to ${\bi n}_i$ from ${\bi n}_i\cdot{\ddot{\bi n}}_i + |{\dot{\bi n}}_i|^2=0$. The right hand side vanishes 
if ${\bi E}_i^{\rm eff}$  is parallel to ${\bi n}_i$. 
From Eqs.(19) and (20) the Hamiltonian ${\cal H}= 
{\cal K}+U $ changes   as 
${d} {\cal H}/dt = - M_z {d}E_{\rm a}/dt$  (without thermostats). 
Thus,  $\cal H$ is conserved  for stationary $E_{\rm a}$.
 
At low $T$,  we have 
$\mu_0 |{\bi E}_i^{\rm eff}|\gg k_{\rm B}T$  for most $i\in 1$, where 
 ${\bi n}_i$ is nearly parallel to $ {\bi E}_i^{\rm eff}$. 
From Eq.(2) we set  
\be  
{\bi E}_i^{\rm eff}={\bi E}_i +{\bi E}_i^{\rm ste},
\en  
where ${\bi E}_i$  is the long-range dipolar part  in Eq.(13) and  
${\bi E}_i^{\rm ste}$  is  the short-range steric  part   from 
the orientation dependence of $U_{\rm LJ}$  in Eq.(3). Some calculations give  
 \be 
{\bi E}_i^{\rm ste} =  -({8\ve {\eta}}/{\mu_0})   
 \sum_{j\neq i} 
({\sigma}_{\alpha\beta}^{12}
/r_{ij}^{14} ) ({\bi n}_i\cdot{{\bi r}}_{ij} ){\bi r}_{ij}.
\en 
where main contributions arise  from  neighbors $j$ with 
  $r_{ij}\ls \sigma_{\alpha\beta}$.  
These neighbor impurities ($j\in 2$) yield  
local  random pinning fields  (see Fig.3(a)).

\subsection{Simulation  method }

We  integrated Eqs.(19) and (20) for $N=N_1+N_2=8000$. We used    the  
 3D   Ewald method on the basis of   $U_{\rm d}$ in Eq.(11)
\cite{Hautman,Klapp,apply,Takae1,Takae2}.
 To realize  crystal, we slowly cooled the system from 
a liquid  above  the  melting temperature ($\sim \epsilon/k_{\rm B}$)
 at density $N/V=0.84\sigma_1^{-3}$. 
In crystal, there is no translational diffusion.
%the imprities play the role of quenched disorder. 
We attached   Nos\'e-Hoover thermostats to  the particles 
in the layer regions $z<2\sigma_1$ and $H-z<2\sigma_1$. 
We fixed the cell  volume at $V=L^2H$   with $H=L=21.2\sigma_1$ 
 mostly, but  we  slightly varied $H$ in time   
 to obtain   the  field-induced strain in Sec.V.

In our system, the dielectric response  strongly 
depends on the dipole moment $\mu_0$ in Eq.(7)\cite{Johnson1}, 
so we present our results for   $\mu_0=0.8$ and    1.6 
in units of $(\epsilon\sigma_1^3)^{1/2}$.  
For example\cite{Johnson0,Johnson1}, if  $\epsilon /k_{\rm B}= 100$ K  and 
$\sigma_1 = 5~{\rm \AA}$,  these values of $\mu_0$ 
 are $1.05$ D  and 2.10 D, respectively.

We measure space and time 
 in units of $\sigma_1$ and 
\be 
t_0= \sigma_1(m/\epsilon)^{1/2}.
\en 
Units of  $T$,  electric potential,  and electric field 
are   $\epsilon/k_{\rm B}$, 
$(\epsilon/\sigma_1)^{1/2}$,  and $(\epsilon/\sigma_1^3)^{1/2}$,  respectively. For  $\epsilon /k_{\rm B}= 100$ K  and 
$\sigma_1 = 5~{\rm \AA}$,  we have 
$(\epsilon/\sigma_1)^{1/2}=  0.16$ V, $(\epsilon/\sigma_1^3)^{1/2}= 
 0.32$ V$/$nm,   
and  $e=18.3(\epsilon\sigma_1)^{1/2}$ (elementary charge).

Because of   heavy    calculations of electrostatics 
 we performed a single simulation run  for each parameter set. 
Then,  $\av{\cdots}$ 
denotes the time average  (not  the ensemble one).
We also do not treat  slow aging processes\cite{aging,aging0,Kob}, 
for which very long simulation time is needed.

%2 
\begin{figure}
\begin{center}
\includegraphics[width=240pt]{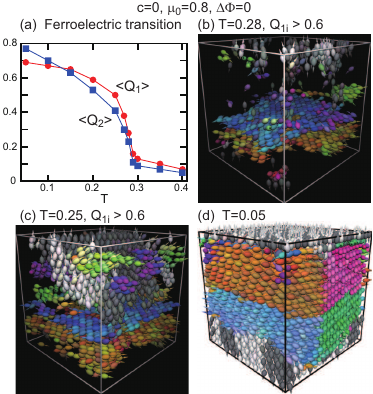}
\caption{ Ferroelectric transition in  one-component 
dipole system (${c=0}$) with  $\mu_0=0.8$  
 without applied field ($\Delta\Phi=0$). 
(a)  Average orientational order 
parameters $\av{Q_1}$ and $\av{Q_2}$ in Eq.(24) 
vs $T$,  where a ferroelectric  transition is steep but  
gradual due to multi-domain states in a film  system 
with fixed width $H= 21.2$.  (b) Ordered regions with $Q_{1i}>0.6$ 
at $T=0.28$  and (c) those at $T=0.25$  in a disordered matrix, where 
the fraction of the ordered regions  
 expands with lowering $T$. 
(d) Rhombohedral polycrystalline state at $T=0.05$.  
}
\end{center}
\end{figure}

\section{ Ferroelectric  transition for $c=0$}

We first examine  a  ferroelectric 
transition in crystal composed of  dipolar 
spheroids with $\eta=1.2$ in Eq.(4) 
without impurities. See  similar simulation by 
 Johnson {\it et al.}\cite{Johnson1} 
for the prolate  case with  aspect ratio  1.25. 

It is convenient to introduce  orientational order parameters 
defined  for each dipole $i$ as 
\be
Q_{\ell i}=\sum_{j\in {\rm neighbor} }
P_\ell({\bi n}_i\cdot{\bi n}_j)/Z_i\quad (\ell=1,2), 
\en 
where     $P_1(x)=x$ and  $P_2(x)=(3x^2-1)/2$. 
We sum over  neighbor  dipoles $j$ 
with $r_{ij}<1.4$, where $Z_i$ is their number. 
Then, $Q_{1i}$ represents the local dipolar order 
and $Q_{2i}$ the local quadrupolar order\cite{Binder,ori}.
These variables will  be used also for  ferroelectric glass with  $c>0$. 

In Fig.2, we examine  the  transition 
 by slowly lowering  $T$ for ${c=0}$,   $\mu_0=0.8$, 
and $\Delta\Phi=0$. In (a), 
we plot the averages  $\av{Q_1}= \sum_{i\in 1}Q_{1i}/N_1$ 
 and $\av{Q_2}= \sum_{i\in 1}Q_{2i}/N_1$. 
Here,  the transition is steep but  gradual 
due to  the finite-size effect imposed by 
the metal walls at $z=0$ and $H$.  In  our system, 
the  spheroidal particles  form a fcc crystal 
%e $z$ axis  along $[111]$ 
in the plastic crystal phase \cite{Binder,ori,EPL}  
in the range  $0.3<T<1$.   
For lower $T$,  a polycrystal with   eight 
rhombohedral variants appears, where the spheroid 
directions $ {\bi n}_i$ are  along $\av{111}$ 
%perpendicular to ${\{}111{\}} $ planes.  
except those near the interfaces.
 
In the transition  range $0.22\ls T\ls 0.30$, 
the system is composed of disordered and ordered regions 
with sharp interfaces. We give snapshots of 
relatively ordered regions with $Q_{1i}>0.6$ 
at (b) $T=0.28$   and (c) $T=0.25$, where we pick  up (b) 10$\%$ 
and (c) $30\%$ of the total dipoles. 
These patterns are stationary in our simulation time intervals. 
In (d), at $T=0.05$, we  give a snapshot of 
polycrystal state  with eight  variants.

The rhombohedral structure 
is characterized by the 
 angles $\pi/2\pm \alpha$ 
of its lozenge faces of a unit cell. At low $T$, we find $\alpha\cong 5^\circ$ 
for $\mu_0=0.8$ 
but   $\alpha\cong 1^\circ$ for $\mu_0=1.6$. 
See Sec.VA for the   reason of this $\mu_0$ dependence.

\section{Ferroelectric  transition for $c>0$ }

\subsection{Role of impurities} 
%3
\begin{figure}
\begin{center}
\includegraphics[width=240pt]{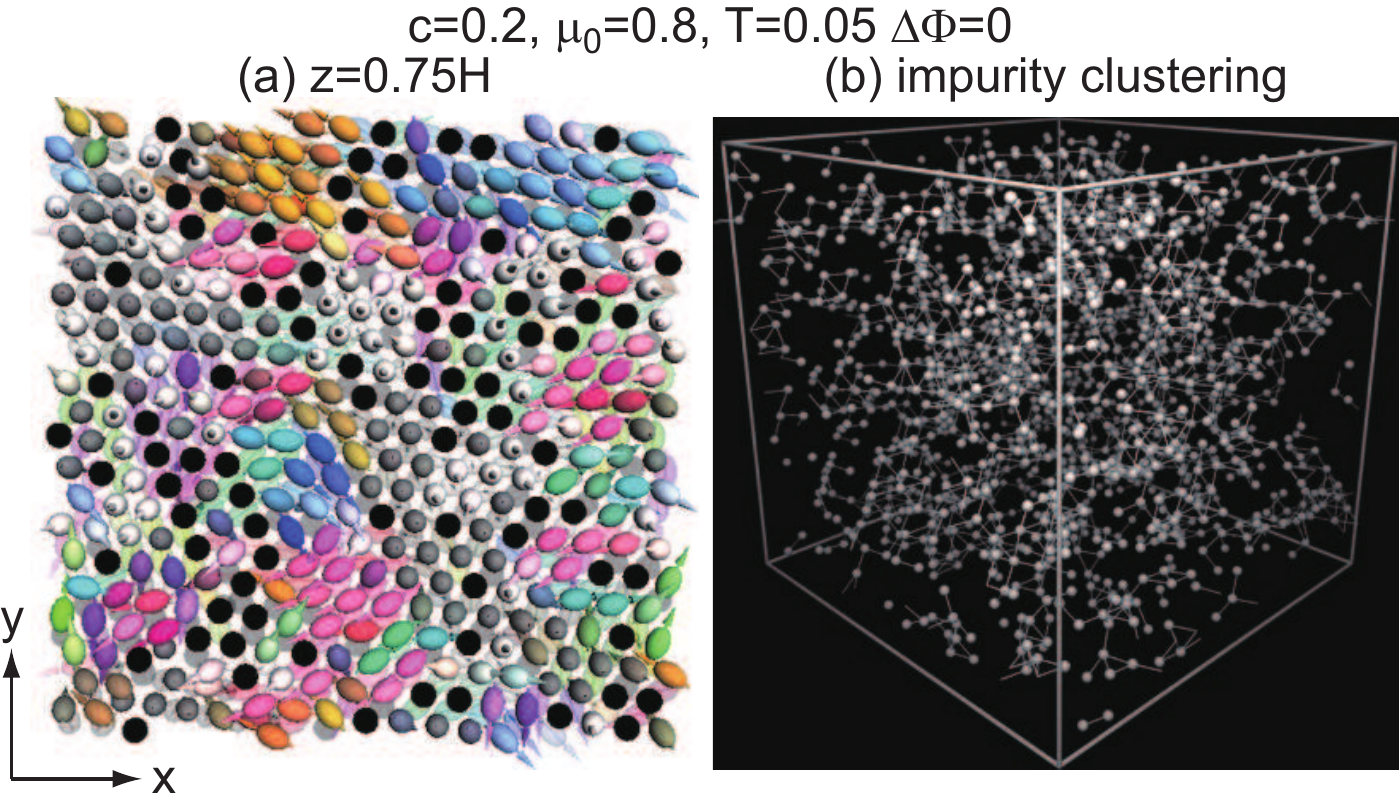}
\caption{(a)  Cross-sectional snapshot on $(111)$ ($xy$-plane) 
at $z=0.75H$ exhibiting planar anchoring of dipoles around 
impurities (black circles).
 (b) Distribution of impurities 
(white spheres)   with diameter 0.22 (real diameter  being 1.1). 
Bonds (white lines) are written between 
  impurity pairs  if $r_{ij}<1.4$. 
 In these panels,  $c=0.2$,  $\mu_0=0.8$,  $T=0.05$, 
 and $\Delta\Phi=0$.}
\end{center}
\end{figure}

The   impurities  hinder 
the spheroid rotations   and   
suppress   long-range orientational  order  
 not affecting  the crystal order.
 In our previous papers  \cite{Takae-ori,EPL}, 
this gave rise to   orientational glass without electrostatic interactions.   
In a mixture of nematogenic  
molecules  and  large spherical particles, surface  anchoring of 
the former  around the latter suppresses  the 
long-range nematic  order \cite{Jun}.

In   Fig.3(a), we  display the 
dipole directions  for c=0.2 on a (111) plane at $z=0.75H$. 
Many of them tend to align in the directions 
parallel to the impurity surfaces or 
 perpendicular to  ${\bi r}_{ij}$ ($j \in 2$), 
 because   $A_{ij}=0$ for  ${\bi n}_i\cdot {\bi r}_{ij}=0$ 
in  Eq.(3).  However,  this  anchoring  is possible   only  
partially,   because the dipoles  are  
on the lattice points and the impurities form clusters.  
This picture resembles those  of PNRs on crystal surfaces  of 
relaxors \cite{Ann,Bokov2,Bokov4}.

In Fig.3(b),  we display all the   impurities in the cell for $c=0.2$,  
where clustering is significant.  
As  guides of eye, we  write  bonds  between 
pairs of  impurities if their distance is smaller than 1.4.
In this bond criterion, we find  large clusters composed of many 
members $(\gs 10)$ including 
a big one percolating through  the cell.    These  clusters 
were pinned   during crystallization, 
so they depend on  the potentials and the cooling rate. 
They strongly influence the shapes of  PNRs (see Figs.9 and 10 also).

Correlated quenched disorder   should  also be relevant  in real systems. 
For relaxors, much effort\cite{Wel,Wel1,Ban} has  been made to 
determine   the  distribution of  the B-site ions 
(Mg$^{2+}$ and Nb$^{5+}$ for PMN)  using 
effective atom-atom interactions, 
while Burton {\it et al.} \cite{Burton1,Burton2,Burton3} 
demonstrated   strong influence of   
 compositional  heterogeneity on the PNRs.

%4 
\begin{figure}
\begin{center}
\includegraphics[width=249pt]{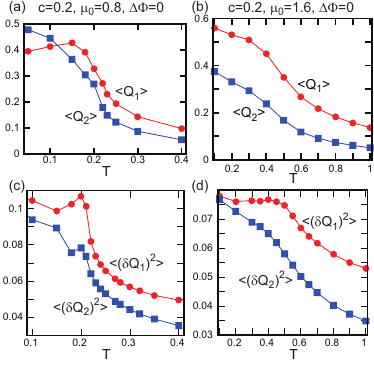}
\caption{ Diffuse ferroelectric transition with impurities for 
$c=0.2$   and  $\Delta\Phi=0$. 
Top:   $\av{Q_1}$ and $\av{Q_2}$ 
vs $T$ for (a)  $\mu_0=0.8$ and (b)
  $\mu_0=1.6$. 
Bottom:   $\av{(\delta Q_1)^2}$ and $\av{(\delta Q_2)^2}$ 
vs $T$ for (c)  $\mu_0=0.8$ and (d)
  $\mu_0=1.6$. 
}
\end{center}
\end{figure}

%5
\begin{figure}[t] 
\begin{center}
\includegraphics[width=240pt]{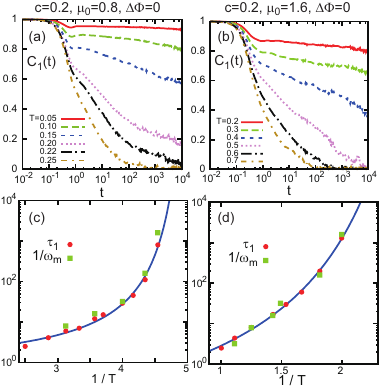}
\caption{ Slow orientational dynamics  with impurities for 
$c=0.2$   and  $\Delta\Phi=0$. 
Top:    Time-correlation function $C_1(t)$ in Eq.(26) 
for (a)  $\mu_0=0.8$ and (b)
  $\mu_0=1.6$ at several temperatures. 
Bottom: Orientational relaxation time $\tau_1(T)$ from Eq.(27) (red circles) 
 and inverse  frequency $1/\omega_{\rm m}(T)$ vs $1/T$ 
from Eq.(29) (green boxes) 
 for (c) $\mu_0=0.8$  and (d) $\mu_0=1.6$, indicating $\tau_1 \omega_{\rm m}
\sim 1$.  As a guide of eye, data of $\tau_1$ are fitted to 
Vogel-Fulcher form (28) (bold line).
}
\end{center}
\end{figure}

\subsection{Diffuse transition toward ferroelectric glass}

The  dipole moment $\mu_0$ 
determines    relative importance of  the dipolar  and 
steric parts,  ${\bi E}_i$ and ${\bi E}_i^{\rm ste}$,  in  
the orientating   field in Eq.(21), since they are proportional 
to $\mu_0$ and $\mu_0^{-1}$, respectively,  
for $E_{\rm a}=0$. 
For example, for   $c=0.2$ and $\Delta\Phi=0$, 
we  average  
 (${|{\bi E}_i|}$, $ {|{\bi E}_i^{\rm ste}|}$)  
over all the dipoles 
 to obtain   ($2.0$, $5.7$) for $\mu_0=0.8$ and  $ T=0.1$ 
and  ($4.6$,  $3.4$)  for $\mu_0=1.6$ and $ T=0.2$.  
Thus,  ${\bi E}_i$  is more important 
for larger  $\mu_0$ in the dipole orientations. 
Here, the amplitude of the local electric field  $|{\bi E}_i|$ 
is mostly of  order  $4\pi\mu_0 n_1/3= 2.8\mu_0$,  
where $n_1=(1- c)N/V$.  This large size of ${\bi E}_i$  is realized   
within  mesoscopic PNRs.

In Fig.4, we examine the transition with  $c=0.2$    and $\Delta\Phi=0$ 
for   the two cases $\mu_0=0.8$ and   $1.6$,
 where the net polarization nearly vanishes. 
At each $T$, we waited for  a  time   $\Delta t\sim 
5\times 10^4$.   In (a) and (b)
 we show gradual $T$ dependence 
of   $\av{Q_\ell}=\sum_{i\in 1}Q_{\ell i}(t_0)/N_1$. 
%where the average over $t_0$ is also taken,
%Since the inner products  ${\bi n}_i\cdot{\bi n}_j$ are used, 
They take appreciable values  in the presence of small PNRs.
In  (c) and (d), we also show their variances, 
\be 
\av{(\delta Q_\ell)^2}= \sum_{i\in 1}  
( Q_{\ell i}(t_0) -\av{Q_\ell} )^2/N_1 \quad (\ell=1,2).
\en   
%which are averaged over $t_0$. 
 The orientation  fluctuations are 
frozen at large sizes 
at low  $T$. We also see that   $\av{Q_1}$ in (a) 
and $\av{(\delta Q_1)^2}$ in (c) exhibit small maxima 
 at low $T$, but they    
  should disappear in the ensemble  averages. 

In Fig.5, we plot  the time-correlation functions $C_1(t)$ 
for one-body angle changes  defined by  
\be 
C_1(t)=\sum_{i\in 1}\av{{\bi n}_i(t_0)\cdot{\bi n}_i(t_0+t)}/N_1, 
\en 
where the average is taken over the initial time $t_0$. 
   In (a) and (b), the angle changes   
slow down with lowering $T$.  We define the  reorientation  time $\tau_1$ by 
\be 
C_1(\tau_1)=0.1,
\en 
where 0.1 is smaller than the usual choice $e^{-1}$ 
since $C_1(t)$ decays considerably 
in   the initial thermal stage for not very low  $T$. 
The PNRs are broken on this timescale. 
In (c) and (d), we display    $\tau_1$ vs $1/T$.
where  $\tau_1$ can well be fitted to the Vogel-Fulcher form \cite{Kob},
\be 
\tau_1= \tau_{10} \exp[D_1T_1/(T-T_1)].
\en  
Here,  $\tau_{10}$, $T_1$,  and $D_1$ are constants 
with $(D_1, T_1)$ being $ (0.89, 0.19)$  for $\mu_0=0.8$  
and $(4.7, 0.32)$ for  $\mu_0=1.6$.

 \subsection{Dielectric permittivity}

%6 
\begin{figure}
\begin{center}
\includegraphics[width=240pt]{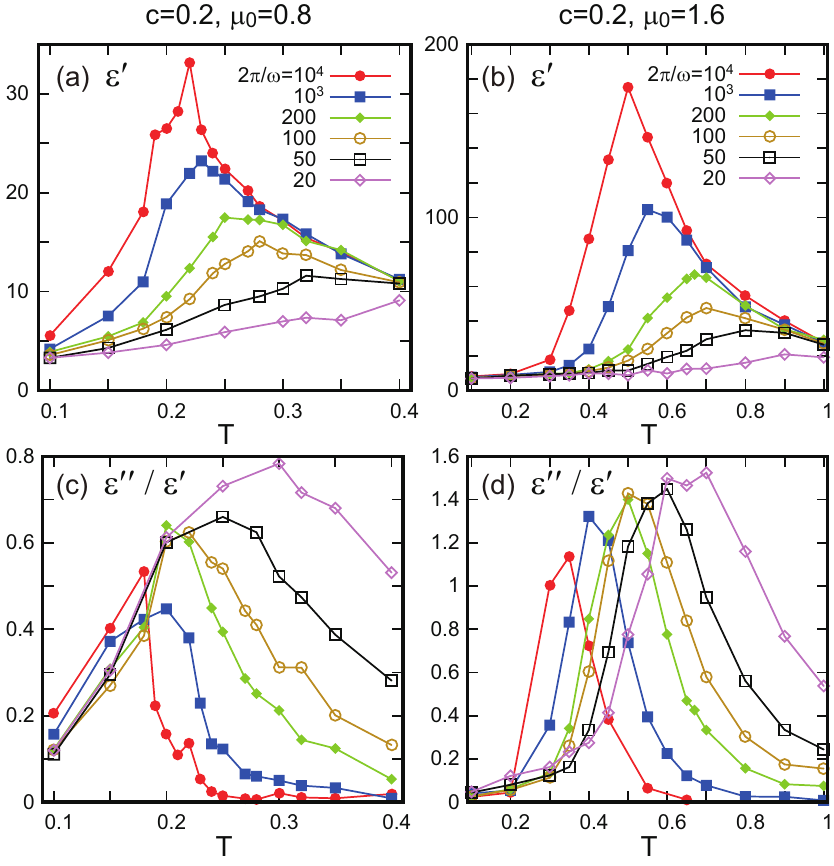}
\caption{ Frequency-dependent dielectric permittivity for 
$2\pi/\omega=10^4, 10^3, 200, 10^2$, 50,  and $20$ 
with  $c=0.2$ and $\Delta\Phi=0$. 
Top:  $\ve'(\omega,T)$  vs $T$ for (a) $\mu_0=0.8$ and (b) $\mu_0=1.6$,  which 
  exhibits   a maximum  $\ve'_m(\omega)$ at $T=T_{\rm m}(\omega)$.  
Bottom:   $\ve''(\omega,T)/\ve'(\omega,T)$  
vs $T$   for (c) $\mu_0=0.8$ 
and  (d) $\mu_0=1.6$.}
\end{center}
\end{figure}

%7
\begin{figure}[t]
\begin{center}
\includegraphics[width=240pt]{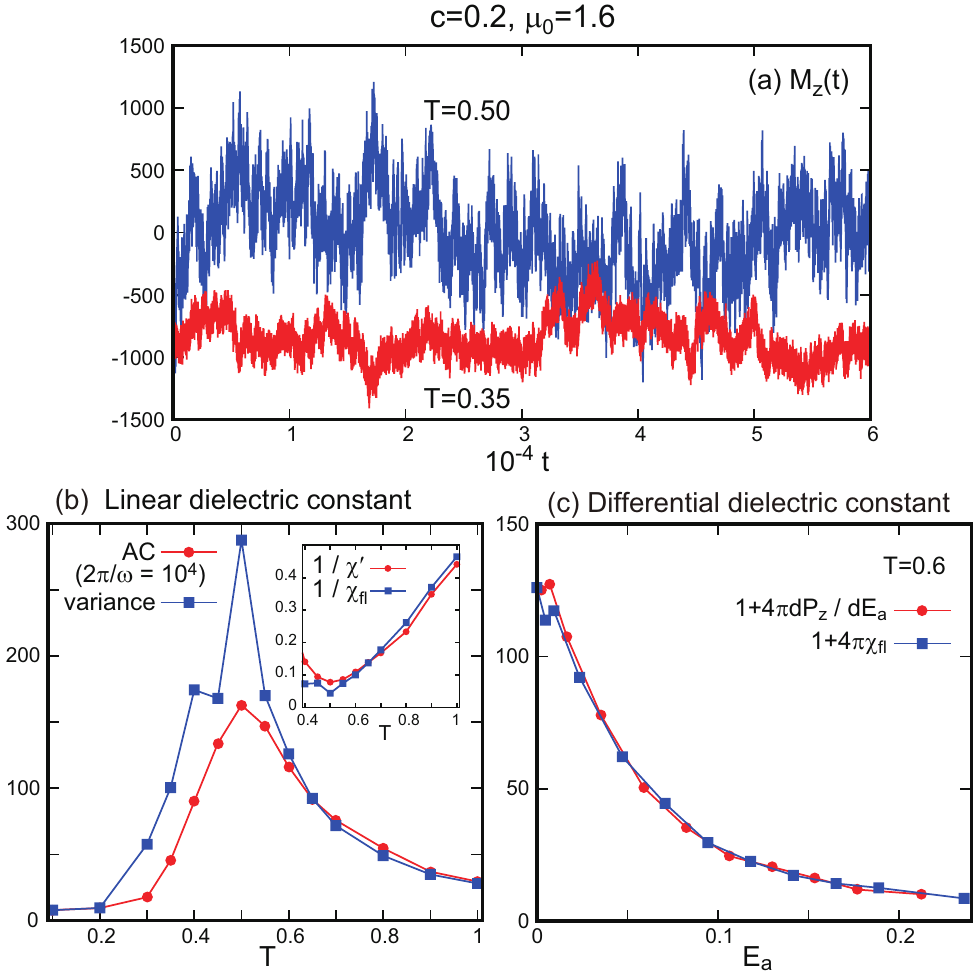}
\caption{Ergodicity and nonergodicity 
for $c=0.2$ and $\mu_0=1.6$. 
 (a) Time evolution of $M_z(t)$ 
at $T=0.5$ (blue) and 0.35 (red) with  $\Delta\Phi=0$. 
(b) $\ve'(\omega,T)=1+4\pi\chi'$ (red) and 
 $\ve_{\rm fl}= 1+4\pi\chi_{\rm fl}$ (blue), where   
$\ve'$ is from ac response at 
$\omega=2\pi \times 10^{-4}$ in Fig.6(b)
and  $\chi_{\rm fl}$ is  the normalized variance 
of $\delta M_z$   in Eq.(30) for $\Delta\Phi=0$. 
Two curves  coincide  for $T>T_{\rm m}(\omega)$, but 
disagree for lower $T$. Shown also are 
  $1/\chi'$ and $1/\chi_{\rm fl}$ (inset), indicating Eq.(31). 
(c) Coincidence of $\ve_{\rm fl}= 1+ 4\pi \chi_{\rm fl}$ 
and $\chi_{\rm dif}= 1+ 4\pi dP_z/dE_{\rm a}$ 
as functions of $E_{\rm a}$ at $T=0.6$. 
 }
\end{center}
\end{figure}

We  next examine the   dielectric permittivity.  
We calculated its real part  
  $\ve'(\omega,T)$  and  imaginary part $\ve''(\omega,T)$ as 
 functions of  $T$ and  the frequency $\omega$  
  by applying  small ac   field in the linear response  regime  (see Appendix B).

In  Fig.6, we show    $\ve'$ 
and  the ratio $\ve''/\ve'$ vs $T$ at several low  frequencies     
 for the two cases  $\mu_0=0.8$ (left) and 1.6 (right).
In (a) and (b),   $\ve'$  increases with decreasing $\omega$ 
and exhibit a broad maximum at a temperature 
$T=T_{\rm m}(\omega)$ for each $\omega$.  
  With decreasing $\omega$,   $T_{\rm m}(\omega)$  decreases (with weaker 
dependence for smaller $\omega$) 
and the peak height  $\ve'_{\rm m}(\omega)=  \ve'(\omega,T_{\rm m}(\omega))$  
increases. For $T>T_{\rm m}$, 
 we have $\omega<\tau_1^{-1}$,  so 
 $\ve'(\omega,T)$ tends to the linear dielectric constant $\ve(T)$. 
However,  for $T<T_{\rm m}$,  $\ve'$   decreases to zero  
with lowering $T$ or increasing $\omega$, 
 where the response of the  PNRs to small ac field decreases.     
On the other hand, $\ve''/\ve'$  
 exhibits a maximum for each $\omega$  
and shifts to a  lower temperature  with lowering $\omega$.    
These    behaviors  characterize    ferroelectric glass
\cite{Cr,Cr1,Smo,Klee,Klee1,Zhang,Cowley,Bokov,Bokov3,Samara,Loidl3,Ho2}. 
%\cite{Cross,Smo,Klee,Ra,Zhang,Cowley,Bokov,Samara,ori,Ho1,Ho2,Mori}. 
Similar behaviors were  found for the frequency-dependent magnetic 
susceptibilities in  spin glass\cite{Mydosh}. 
 Furthermore,    in Appendix B, we will present   
analysis of  $\ve'$ and $ \ve''$ for  $\omega\gs  \tau_1^{-1}$ 
at relatively high $T$ on the basis of the linear response theory\cite{Kubo}.

We write  the  inverse relation of $T=T_{\rm m}(\omega)$  as  
\be 
\omega=T_{\rm m}^{-1}(T)=\omega_{\rm m}(T),
\en 
leading to  $\ve'(\omega_{\rm m},T)= \ve_{\rm m}'(\omega_{\rm m})$. Here,   
  $T/T_{\rm m}>1$ ($<1$) holds for  
$\omega/\omega_{\rm m}<1$ ($>1$). 
In (c) and (d)  of Fig.5, we compare the inverse  $1/\omega_{\rm m}(T)$ 
and   $\tau_1(T)$ in  Eq.(27) 
 for $\mu_0=0.8$  and 1.6. We  find  $ \omega_{\rm m} \sim \tau_1^{-1}$.
Thus,  $\omega_{\rm m}(T)$ 
represents a characteristic frequency  of the dipole reorientations.
Previously, for relaxors and spin glasses,  
Stringer {\it et al.}\cite{Ra} nicely fitted 
$1/\omega_{\rm m}(T)$ to the Vogel-Fulcher form, which is in accord with 
(c) and (d) of Fig.5.

Our   system is  ergodic at relatively high $T$,   
but becomes   nonergodic  as  $T$ is lowered. 
The boundary between these two regimes weakly depends on 
the observation time. In Fig.7(a),   $M_z(t)$ evolves 
on   a wide  range of time scales in a time interval with width $6\times 10^4$ 
for $c=0.2$, $\Delta\Phi=0$, and $\mu_0=1.6$.
At $T=0.5$, its time average becomes  small, but its  fluctuations 
are large. In contrast, at $T=0.35$,  it remains  
negative around $ -800 = - 14 (Vk_{\rm B}T)^{1/2}$, 
on which smaller thermal fluctuations with faster  time scales  
 are superimposed.   
Note that   the ensemble  average of $M_z$  should 
vanish  at any $T$ for $\Delta\Phi=0$.  

For a single simulation run, 
we consider the time average of the normalized polarization variance, 
written as $\chi_{\rm fl}$. To avoid confusion, we define  it explicitly as 
\be
Vk_{\rm B}T \chi_{\rm fl}={{\av{(\delta M_z)^2}}_{\rm time }}=
 {\av{M_z^2}}_{\rm time } 
- {\av{M_z}}_{\rm time}^2.  
\en 
We set 
 ${\av{{\cal A}}}_{\rm time } = \int_{t_1}^{t_2} dt 
{\cal A}(t) /\Delta t$  
with  $\Delta t= t_2-t_1 (\sim 5\times 10^4$ here) for 
any time-dependent variable ${\cal A}(t)$. This averaging procedure has already been taken for the quantities in Figs.4 and 5.  In the nonergodic  $T$ range, 
${\av{M_z}}_{\rm time}$ remains  nonvanishing 
 even for $\Delta\Phi=0$, while $\chi_{\rm fl}$ arises from the 
 (thermal)  dynamical  fluctuations and 
tends to zero as $T\to 0$.
   In Fig.7(b), we plot   numerical  results of   
 $\ve_{\rm fl}= 1+4\pi   \chi_{\rm fl}$ for $\Delta\Phi=0$ 
and $\ve'=1+4\pi \chi'$ at   $\omega= 2\pi\times 10^{-4}$ 
as functions of $T$.  
  These two curves nearly  coincide  
for $T>T_{\rm m}$ yielding $\ve(T)$,  
 but    $\ve_{\rm fl}$  is considerably larger than $\ve'$  for $T<T_{\rm m}$. 
In their simulation,  Burton {\it et al.}\cite{Burton1,Burton2} 
calculated  a dielectric constant  from  polarization fluctuations, 
which corresponds to  $\ve_{\rm fl}$ in our case.

In Fig.7(b),       $\ve_{\rm fl}$ and $\ve'$ 
 steeply grow as $T\to T_{\rm m}$.  From  
the curves of $1/\chi'$ 
and $1/\chi_{\rm fl}$ in  its  inset, 
$\chi'= (\ve'-1)/4\pi$  and  $\chi_{\rm fl}$   can fairly be 
 fitted to the Curie-Weiss  form,  
\be 
\chi'\cong \chi_{\rm fl}   \cong A_0/(T-T_0),
\en 
  with  $A_0 \cong 1.2$ and $T_0\cong 0.47 (\cong T_{\rm m}$ 
at  $\omega=2\pi\times 10^{-4}$)  
 for $T\gs 0. 55$. 
At $T=0.5$, however, we find  $\chi'\cong 13$ and $\chi_{\rm fl}\cong 23$. 
In experiments, the behavior (31) was found for 
orientational glass\cite{ori}, but 
 a marked deviation  was detected  close to $T_{\rm m}$  
for relaxors\cite{Samara,Bokov,Cr1}.   
Thus, if  $T$ is somewhat above $T_0$, 
our   polarization fluctuations  
   resemble the critical fluctuations 
in  systems near their critical point \cite{Cowley,Samara}. 
In our disordered  system, 
these {\it near-critical}  fluctuations are  
slowed down and eventually frozen    as  $T$ is further lowered, as in relaxors. This  can  also be seen  in (c) and (d) of   Fig.4. 
Furthermore, for $T\ls T_0$,  there is a tendency of interface 
formation between adjacent PNRs for $c\ls 0.2$,  
which will be discussed in future. 

For relaxors,  Stock {\it et al}.\cite{Stock} 
 divided the scattering intensity into frozen 
 and  dynamic parts, where the former (latter) increases (decreases) 
with lowering $T$.  
Similar arguments of nonergodicity were  made for polymer 
gels\cite{Pusey,Matsuo}, where  
 the   fluctuations  of the polymer density 
consist of frozen   and dynamic  parts. Moreover, 
 if gelation takes place 
in a polymer solution   close to its   criticality, 
the  critical  concentration 
fluctuations  are pinned at the  network formation 
\cite{Matsuo,Onukibook}.

We next  confirm Eq.(17)  by  increasing $E_{\rm a}$ at $T=0.6$ 
with  $c=0.2$  and $\mu_0=1.6$, where the observation time 
is much longer than  $\tau_1 \sim 60$. In Fig.7(c),  
we compare the differential formula 
$\ve_{\rm dif}= 1+ 4\pi dP_z/dE_{\rm a}$ 
and the  fluctuation formula  
$\ve_{\rm fl}= 1+ 4\pi\chi_{\rm fl}$ 
 for the field-dependent  dielectric constant. 
The  former is calculated from  the data in    Fig.12(a) 
and the latter from Eq.(30), where  
  these  two curves  are surely very close. 
At this $T$, the polarization fluctuations 
are   suppressed  with increasing $E_{\rm a}$.

%8
\begin{figure}
\begin{center}
\includegraphics[width=249pt]{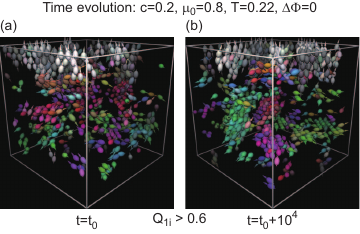}
\caption{  Time evolution of 
%orientational heterogeneities 
PNRs  in the ergodic regime, where  $T=0.22$,  $c=0.2$,  and 
$\mu_0=0.8$  (see (a) and (c) of Fig.5). 
 Displayed  are snapshots of  
dipoles with $Q_{1i}>0.6$  at (a) $t=t_0$  and 
(b)  $t=t_0+10^4$  in the same simulation run, 
which amount to  $14\%$ of the total dipoles.  
These patterns are different, so they  have lifetimes shorter than  $10^4$. 
}
\end{center}
\end{figure}

\subsection{Polar nanoregions in  diffuse transition} 

In our diffuse transition,    the PNRs  are relatively 
ordered regions  consisting  of 
 aligned clusters  enclosed by impurities.    
At relatively high $T$, they  have  finite lifetimes 
(within observation times)\cite{Bokov,Bl}. 
This feature is illustrated  in  two snapshots in   Fig.8, 
 which  were  taken at two  times  
separated by $10^4$ in the same simulation run. 
They display the dipoles with $Q_{1i}>0.6$ for  
$T=0.22$,   $c=0.2$,  and $\mu_0=0.8$. 
These  two  patterns   are very different, so   
their   lifetime is   shorter than $10^4$. In fact, 
   $\tau_1$ is  of order $10^3$  at $T=0.22$ in Fig.5(c).

%9 
\begin{figure}
\begin{center}
\includegraphics[width=249pt]{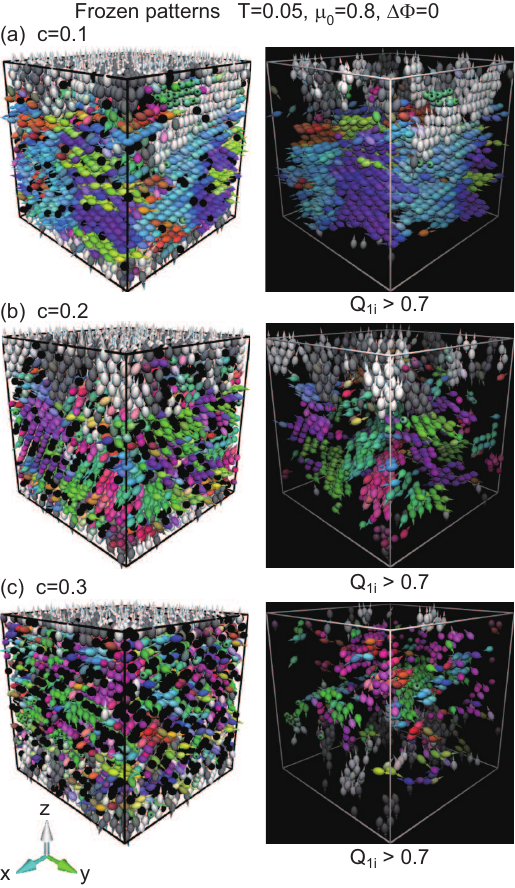}
\caption{Frozen PNRs at $T=0.05$ for (a) $c=0.1$ (top), 
(b) $c=0.2$ (middle), and (c) $c=0.3$ (bottom) 
 with $\mu_0=0.8$ and $\Delta\Phi=0$. 
Left: Dipoles  (in color) and impurities (in black) on the 
boundaries ($x,y,{\rm or}~z=L)$.
Right: Dipoles  with $Q_{1i}>0.7$ forming PNRs, whose 
typical sizes decrease with increasing $c$.
}
\end{center}
\end{figure}

%10
\begin{figure}[tbp]
\begin{center}
\includegraphics[width=210pt]{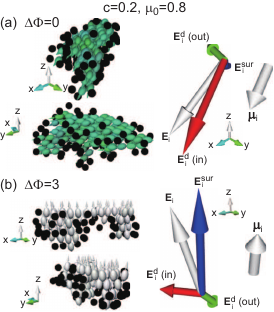}
\caption{  Left: Frozen PNRs   
 surrounded by impurities (black spheres) viewed from two directions 
for (a)  $\Delta\Phi=0$ at  the cell center  
and for (b) $\Delta\Phi=3$ near the top wall, 
where $c=0.2$, $\mu_0=0.8$, and $T=0.05$. 
Right: Local electric field ${\bi E}_i$ in Eq.(13) and 
dipole moment ${\bi \mu}_i$ of  a typical dipole  $i$ 
within the left PNRs. Here, ${\bi E}_i$ consists of  
the field from the surface charges 
${\bi E}_i^{\rm sur}$ and those from the dipoles 
inside and outside the PNR,  ${\bi E}_i^{\rm d}$(in) 
and ${\bi E}_i^{\rm d}$(out). In (a), 
${\bi E}_i\cong {\bi E}_i^{\rm d}({\rm in})$ 
and ${\bi E}_i^{\rm sur}$ is small. In (b), 
${\bi E}_i\cong {\bi E}_i^{\rm sur}$.
}
\end{center}
\end{figure}

%11
\begin{figure}
\begin{center}
\includegraphics[width=240pt]{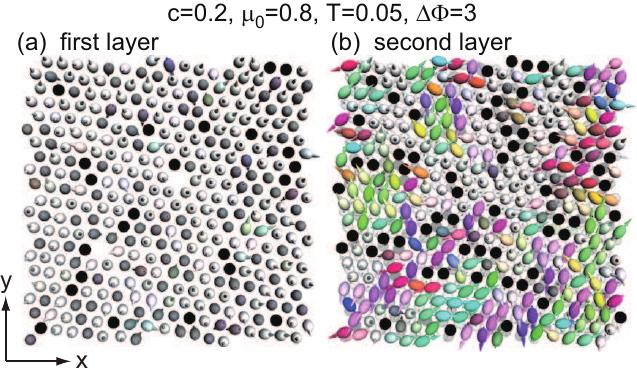}
\caption{ Orientations of  dipoles on $(111)$ planes  near the bottom metal 
wall  for $c=0.2$,  $\mu_0=0.8$, and $T=0.05$ with  
    $\Delta\Phi=3$ ($E_{\rm a}=0.14$). 
Displayed  are cross-sectional snapshots 
in (a) the first  layer ($0<z_i<1$)  and 
 (b) the second layer ($1<z_i<2$), where black 
circles represent impurities. 
In  (a), the fraction of the dipoles 
parallel (antiparallel) to  the $z$ axis is $65\%$ ($35\%$). 
In (b), the oblique orientations appear.}
\end{center}
\end{figure}

The PNRs are frozen  with  lowering  $T $. In  Fig.9, we give  examples 
 for $c=0.1$, 0.2, and 0.3 
with  $T=0.05$, $\mu_0=0.8$,  and $\Delta\Phi=0$.   The left panels display   
 the particles on the boundaries ($x,y,{\rm or}~ z= L$), 
while the right ones  the relatively ordered  dipoles with $Q_{1i}>0.7$. 
 The  dipoles depicted  in the latter 
 amount   to 37, 20, and 13$\%$ of the total dipoles from above.  
 For  $c=0.1$, we can see well-defined 
ordered  domains consisting of eight  variants, whose interfaces are 
trapped at impurities\cite{EPL,Takae-ori}  
(see  Fig.10(a)). These domains  are broken up into 
smaller PNRs with increasing $c$.  For $c=0.2$, 
the   PNRs mostly take compressed,  plate-like shapes 
under the constraint of the spatially correlated  impurities 
(see Fig.10 also). 
For  $c=0.3$, the dipole orientations 
are highly frustrated on the particle scale without 
 well-defined interfaces.

To be  quantitative, we define   PNRs   as follows. In each  PNR, 
any member $i$ satisfies $Q_{1i}>0.7$ and $r_{ij}<1.4$ 
for some $j$ within the same  PNR. In Fig.9, 
the  dipole number in a  PNR is  
 $2200, 270$, and $56$ on the average from above. 
Thus, the connectivity of the PNRs 
sensitively depends on $c$. In the following,
 we treat the case $c=0.2$. 

\subsection{Single polar nanoregion and local electric field}

We visualize individual PNRs frozen at low $T$. 
When the system is composed of PNRs, the local electric field 
 ${\bi E}_i$  in Eq.(13) arises mainly from  the dipoles 
within the same PNR in the bulk. Its  amplitude is of order 
$4\pi\mu_0 n_1/3\cong 2.8\mu_0$ 
for not very large $\Delta\Phi$.

In Fig.10, we pick up (a) a single PNR for 
 $\Delta\Phi=0$ at the cell center and (b) another one for 
$\Delta\Phi=3$ ($E_{\rm a}=0.14$) near the upper wall, 
where  $c=0.2$ and $T=0.05$.
 We depict  the  impurities  whose distance to some 
dipole in the PNR  is shorter than 1.4. We find the numbers of 
the  constituent dipoles and impurities as  (a)  $(120,70)$   
 and (b)  $(70,50)$ using the definition of PNRs 
in Sec.IVD. Here, the dipoles tend to be  parallel to 
the impurity surfaces, as discussed in Sec.IVA,  
and almost all the impurities are on the PNR 
boundaries, resulting in plate-like   PNRs.

In  Fig.10 (right), we choose a typical dipole in the  PNR  interior 
(not in contact with the impurities) and display  its  ${\bi{E}}_i$ and 
 ${\bi \mu}_i$,  where     they are nearly  parallel.  
Here, we divide the dipolar part of    ${\bi{E}}_i$ 
 into    the contributions 
from those  inside and outside the PNR, 
  written as  ${\bi E}_i^{\rm d}(\rm{in})$ and 
${\bi E}_i^{\rm d}(\rm{out})$. Then, 
Eq.(A9) in Appendix A gives 
%${\bi{E}}_i$ is then decomposed  as  
\be 
{\bi E}_i= {\bi E}_i^{\rm d}({\rm in})+{\bi E}_i^{\rm d}({\rm out})+ {\bi E}_i^{\rm sur},
\en  
where the last term arises from  the surface charges. 
 In (a), we find  ${\bi E}_i\cong {\bi E}_i^{\rm d}(\rm{in})$, 
which occurs mostly  for the dipoles in the  interior of  PNRs 
in the bulk. In (b), on the other hand,  we find 
${\bi E}_i\cong  {\bi E}_i^{\rm sur}$, 
where $|{\bi E}_i^{\rm sur}|$ is of the same order as 
$ 4\pi{\bar\sigma}_0=2.4$ and is much larger than $ E_{\rm a}=0.14$. Here,  
${\bar\sigma}_0$ is the mean  surface charge density at $z=0$.   
For example, if we set $\epsilon /k_{\rm B}= 100$ K  and 
$\sigma_1 = 5~{\rm \AA}$, 
we have ${\bar\sigma}_0\sim 0.01 e/\sigma_1^2$. 
for    $\Delta\Phi=3$.

\subsection{Orientation near metal surface }

As  can be seen in Figs.9 and 10(b), 
the dipoles next to the  walls  
are parallel or antiparallel to  the $z$  axis 
 (along $[111]$), whose distances 
from the walls are about 0.5. 
This is  due to their interaction with the image dipoles in the walls   
(see Appendix A)\cite{Takae1,Takae2,apply}. 
 For $\Delta\Phi=0$, these two orientations 
appear equally on the average  due to  the 
 top-tail symmetry of our  spheroidal dipoles. 
For $\Delta\Phi \neq 0$,   one  of them  is more preferred than the other.
 In Fig.11, we show the particles in the first and second $(111)$ layers 
 in  applied field  with $\Delta\Phi=3$, where   
$c=0.2$,  $\mu_0=0.8$, and $T=0.05$. The   parallel and 
antiparallel orientations  appear  in the first layer, 
but the other oblique ones also appear in the second layer. 
We shall see that  the corresponding  surface charge density 
$\sigma_0(x,y)$  at $z=0$  is highly heterogeneous in Fig.17 in   Appendix A.

In our crystal case,    the $\ell$-th layer is   
given by   $\ell-1<z< \ell$, since 
 the separation between  adjacent $(111)$ planes   
is close to  1.  Here, we consider the average of 
$n_{zi}=\cos\theta_i$ over the dipoles in the  $\ell$-th layer 
and write it as $\av{n_z}_\ell$. 
In Fig.11, it is    0.30 for $\ell=1$,   
  $0.34$ for $\ell=2$,  and    $\av{n_z}_{\rm b}=
0.33$ for $\ell\gg 1$. 
These values  are close, so the surface effect  
on the polarization is  weak in this case of our model. 
The excess potential drop 
near the bottom  wall  is  given by 
$4\pi \mu_0 n_1 \sigma_1 
\sum_{\ell} [\av{n_z}_{\rm b}-\av{n_z}_\ell]\cong 0.13$, 
 which  is much smaller than the total drop  $\Delta\Phi=3$.  
In contrast, for highly polar liquids 
such as  water \cite{Hautman,Takae2,Willard}, 
a  significant  potential drop appears in a microscopic (Stern) layer  
on a solid surface  even without ion adsorption.

\section{ Polarization and strain  in applied electric  field}
\subsection{Applying  electric field along $[111]$ at fixed stress}

In this section, we give results of  cyclic changes  of  
$\Delta\Phi=H E_{\rm a}$  for $c=0.2$ and   
$\mu_0=1.6$. We  also calculated the response 
with  $\mu_0=0.8$ (not shown here).
For  these two $\mu_0$ values, 
the characteristic features are  nearly the same, 
but the  response sizes  are  very different. 
That is, the dielectric response   for  $\mu_0=1.6$  is larger  than that for  
$\mu_0=0.8$ by one order of magnitude as in Fig.6,  
while   the  field-induced strain 
  for $\mu_0=1.6$ is about $20\%$ of that for  $\mu_0=0.8$.
See the last paragraph of Sec.III for the rhombohedral angles 
in our simulation. Using   a barostat,  
 we fixed the $zz$ component of the average  stress 
and varied    the cell width  $H(t)$ 
to  calculate  the  field-induced 
strain. The lateral cell length was fixed at   $L$.

In our model,   dipole   alignment along $[111]$ 
  yields both  
steric repulsion and dipolar attraction between adjacent $(111)$ planes. 
Their relative importance  depends on $\mu_0$. If the former 
is larger (smaller) than the latter, an expansion 
(a shrinkage) of the cell width $H$ occurs   for 
$\Delta\Phi\neq 0$. 
Note that the  dipolar interaction between  two dipoles 
at ${\bi r}_i$ and ${\bi r}_j$ 
aligned along  the $z$ axis 
is attractive (repulsive) if the angle between their 
relative vector ${\bi r}_i-{\bi r}_j$ 
and the $z$ axis is smaller (larger) than $\cos^{-1}(1/\sqrt{3})$.

We increased  $\Delta \Phi(t)$  from 0 to 10, 
decreased to $-10$, and then increased again to 10 at fixed $T$ 
without dislocation formation.  
The changing rate ${\dot\Phi}= d(\Delta\Phi)/dt$ 
was $\pm 1.5\times 10^{-3}$ $\epsilon/(\sigma_1^{3}m)^{1/2}$. 
The average pressure along the $z$ axis 
was 0.4 at $T=0.2$ and 3.6  at $T=0.6$ 
 in units of $\epsilon/\sigma_1^3$,  
while the lateral one 
increased by 0.8  for a change of $\Delta\Phi$ from 0 to 10. 
The  $H(t)$ changed   from  $H(0)= L$  at most by $2\%$.  
We calculated  the average 
polarization and strain for $t>0$ given by 
\be 
{P}_z(t) = {M_z}(t) /[L^2H(t)], \quad 
\bar{\gamma} (t) = H(t)/L -1.
\en 
We also calculated the mean 
surface charge density ${\bar\sigma}_0$ 
at $z=0$  to confirm 
  Eq.(A5) in Appendix A (see  Fig.12).

%12
\begin{figure}
\begin{center}
\includegraphics[width=240pt]{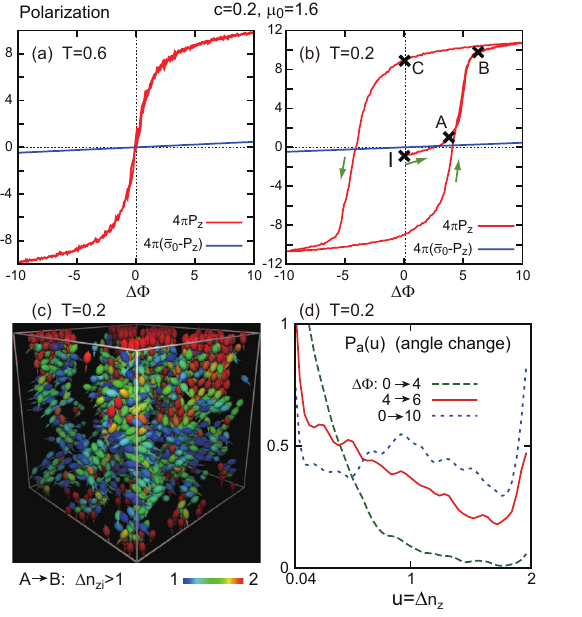}
\caption{Polarization $P_z$  to  cyclic 
applied electric field $E_{\rm a}(t) \cong \Delta\Phi(t)/21$ 
for $c=0.2$ and $\mu_0=1.6$. 
 Top:    $4\pi{{P}}_z$ vs $\Delta\Phi(t)$ at (a) $T=0.6$ 
and (b) $T=0.2$. Straight lines in (a) and (b) (in blue) represent 
$4\pi({\bar\sigma}_0-{ P}_z)$, which coincide with $E_{\rm a}$. 
Bottom: (c) Snapshot of the dipoles with large angle changes 
$\Delta n_{zi}>1$ between two points A and B 
($\Delta\Phi: 4\to 6$)  in (b), where 
$n_{zi}$ is the $z$ component of 
${\bi n}_i$.  Colors represent  $\Delta n_{zi} $ according to the color bar. 
(d) Distribution   $P_{\rm a}(u)$   for $u= \Delta n_{zi}=
n_{zi}(t_1)-n_{zi}(t_0)$ 
between two points  in the cycle in (b).
Small angle changes are dominant  
in the initial interval  ($\Delta\Phi: 0\to 4$) (green line), 
but large angle changes are dominant in  the subsequent one 
($\Delta\Phi: 4\to 6$) (red line). 
}
\end{center}
\end{figure}

%13
\begin{figure}[tbp]
\begin{center}
\includegraphics[width=240pt]{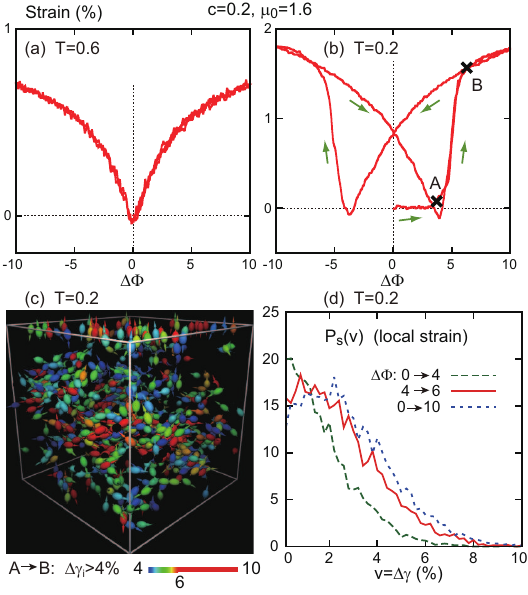}
\caption{ Average strain $\bar\gamma$  to   
applied electric field   for $c=0.2$ and $\mu_0=1.6$. 
 Top: ${\bar \gamma}$ vs $\Delta\Phi(t)$ at (a) $T=0.6$ 
and (b) $T=0.2$.  (c)  Snapshot of the dipoles whose 
changes in the local strain $\Delta \gamma_i $ in Eq.(35) 
 exceed $0.04$ between two points A and B in (b). 
Colors represent  $\Delta \gamma_i $ according to the color bar. 
(d)  Distribution 
 $P_{\rm s}(v)$  for strain changes $v=\Delta\gamma_i=\gamma_i(t_1)
-\gamma_i(t_0)$ between two times in (b). 
It is narrower  for  the initial interval  ($\Delta\Phi: 0\to 4$) (green line) 
than for  the subsequent  one ($\Delta\Phi: 4\to 6$)  (red line).
}
\end{center}
\end{figure}

\subsection{Polarization response}

In Fig.12, we plot $4\pi{P}_z$ vs $\Delta\Phi$ 
for  (a) $T=0.6$ and (b) $T=0.2$. 
At $\Delta\Phi=10$, we have  $P_z\sim  10$  and 
$\av{n_z}= P_z/n_1\mu_0 \sim  0.75$   in (a) and (b), where 
$n_{zi}$ is the $z$ component of ${\bi n}_i$ 
 and  $n_1=N_1/V$ is $ 0.672$. In (a), 
 there is no hysteresis and the initial slope 
 yields  $\ve_{\rm dif} =1+4\pi dP_z/dE_{\rm a}\cong 153$.  In  (b),  
 marked hysteresis appears, where $\ve_{\rm dif}\cong 13$  
    initially at point I, but is 
about $150$ between  two points A and B ($\Delta\Phi: 4\to 6$).
Here, the initial $P_z$ at I is slightly negative as  a frozen 
fluctuation (see the curve at $T=0.35$ in 
Fig.7(a)). At point C on the vertical axis 
we have  a remnant  polarization $P_{\rm R}=0.72$ 
%and  surface charges $ \pm P_{\rm R}$  at $z=0, H$ 
 with  $\Delta\Phi=0$. 
For any $T$,   $\ve_{\rm dif}$  from the initial 
slope at  $E_{\rm a}=0$ 
 nearly coincides with $\ve'$ at $\omega=2\pi \times 10^{-4}$  
in Fig.7(b) (equal to  50 at 
$T=0.4$ and to 10 at $ T=0.1$).  The   curves in (a) and (b) 
closely   resemble those in various ferroelectric systems
\cite{Bl,Mori,Furukawa,Zhang,piezo}.

The field-induced change from A to B 
in  (b) is very steep with large $\chi_{\rm dif}$. 
In (c), we thus  display  the dipoles 
 with large angle changes: 
 $\Delta n_{zi}= n_{zi}(t_B)-n_{zi}(t_A)>1$, where 
$t$ is $t_A$ at A and $t_B$ at B. Collective  reorientations are marked 
in this time interval. In (d),  for three  intervals, 
 we plot the distribution function 
$P_{\rm a}(u)= \sum_{i \in 1}{\delta(u  -\Delta n_{zi})}/N_1$ 
for  $\Delta n_{zi}= n_{zi}(t_1)-n_{zi}(t_0)$, 
where we  use  an appropriately smoothed  $\delta$-function.  
Small angle changes are dominant 
in the first interval $t_I<t<t_A$ (where $t=t_I$ at I), but  
 large angle changes are dominant 
in the next interval  $t_A<t<t_B$. 
%In addition, we remark  that large-angle orientation changes 
%are conspicuous even at $T=0.6$ for small $\Delta\Phi$  in (a).  

 In (b), the initial point I (at $t=t_I$) of the 
cycle represents  an   arrested  state with 
frozen fluctuations  realized by zero-field cooling. 
 It can no longer be reached once a large field is applied. 
The corresponding states  have 
 been realized  in many systems   (see Sec.VI).
%This {\it relaxor state}  is unique  to ferroelectric glass 
%with frozen  PNRs\cite{Bokov,Cross,Kleemann,Samara}. 
In the two states at I and  C, 
the polarization directions are very different, 
but the values of   the potential energy $U$ in Eq.(2) 
are close as  $-8.07N\epsilon$ at I and   $-8.04 N\epsilon$ at  C. 
We can also  see that   the quadrupolar order 
parameters $Q_{2i}(t) $ in Eq.(24) do not change much for most  $i$ 
during the cycle despite large changes in ${\bi n}_i$. 
For example, the mean square difference 
$\sum_i [ Q_{2i}(t_I)-  Q_{2i}(t)]^2/N_1$ 
for time interval $[t_I, t]$ is  $0.047$, $0.103$, and 
$0.086$  at $t=t_A$, $t_B$,
and $t_C$ (which are the times at A, B, and C), respectively, 
where the variance 
$\sum_{i\in 1} [\delta Q_{2i}(t)]^2/N_1$ for $\delta Q_{2i}= 
Q_{2i}- \av{Q_2}$  remains  of order 0.08 (see Eq.(25) and  Fig.4d)

Between  A and B in (b), 
%$\chi_{\rm dif}= dP_z/dE_{\rm a}$ is large and   
we found an increase in    
the   polarization variance $\av{(\delta M_z)^2}(t)$ $(t_A\ls t\ls t_B$). 
   For relaxors, Xu {\it et al.} \cite{Shi2} 
detected an increase in 
 the diffuse scattering in the field range 
with large $\chi_{\rm dif}$. 
We should  then  examine  the scattering 
amplitude between A and B.   In addition,  when $\Delta\Phi$ was  held fixed 
 at 4.0 (at A), we observed  slow reorientations leading to 
coarsening of PNRs\cite{aging,aging0}. 
These effects will   be studied  in future.

\subsection{Field-induced strain }

%14
\begin{figure}
\begin{center}
\includegraphics[width=240pt]{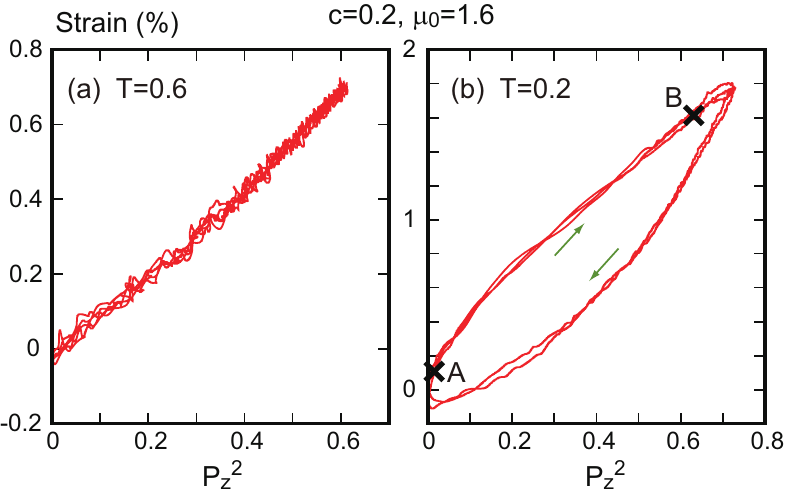}
\caption{Average strain  ${\bar \gamma} $ vs  
${P}_z^2$  for $c=0.2$ and $\mu_0=1.6$. (a) At $T=0.6$, electrostriction 
relation $ {\bar \gamma}\propto P_z^2$ 
holds. (b) At $T=0.2$, a closed loop appears, 
where points A and B corresponds to A and B in (b) of 
Figs.12 and 13. 
}
\end{center}
\end{figure}

In  our model,  the heterogeneity in the strain 
is marked because of dilation of PNRs along $\av{111}$, 
though it is milder than that of the polarization.
 To illustrate  this effect,   we define a 
 local strain $\gamma_i$ along the $z$ axis for each particle $i$ (including the impurities)   by 
\be 
\gamma_i=\sum_{j} | z_j-z_i|/({W}_i a_{111} )-1,
\en  
where the summation is  over other $j$ with $r_{ij}<1.4$ and $|z_j-z_i|>0.2$, 
 ${W}_i$ is the number of these neighbors, 
and $a_{111}(\cong  1.0)$ 
is the average spacing between two consecutive 
$(111)$ planes. From these  conditions, 
the $(111)$  plane containing   $j$ is adjacent  to that containing $i$. 
The  particle average $\sum_i\gamma_i/N$ 
nearly coincides with 
$\bar \gamma$ in Eq.(33).

In Fig.13, we  plot $\bar\gamma$ vs $\Delta\Phi$ with $\mu_0=1.6$ 
in the same simulation run as in Fig.12. 
We  find   (a)  a  cusp curve at   $T=0.6$ and (b) 
a butterfly-like curve  at  $T=0.2$. In (b), 
$\bar\gamma$ becomes slightly negative at $\Delta\Phi\cong \pm 4$. 
These two  curves  
resemble those in the previous experiments 
\cite{Furu1,Zhang,piezo}.
In (c), we pick up the particles with large local strain changes 
$\Delta \gamma_{i}= \gamma_{i}(B)-\gamma_{i}(A)>0.04$ 
between two points A and B  at $T=0.2$ in (b), where $\bar\gamma$ 
is $0.016$ at B.   We define  the distribution function, 
$
P_{\rm s}(v)= \sum_i {\delta(v-\Delta\gamma_i)}/N
$  
for strain changes $\Delta\gamma_i=\gamma_i(t_1)
-\gamma_i(t_0)$ between two times in (b).   In (d), 
it is narrower  for  the initial interval  ($\Delta\Phi: 0\to 4$) 
than for  the subsequent  one ($\Delta\Phi: 4\to 6$).

The shapes of our   dipolar spheroids 
 are  centrosymmetric,   leading to  the electrostriction relation,   
\be 
\bar{\gamma}\cong  C_{\rm es} P_z^2, 
\en  
  at relatively high $T$.  In   Fig.14,  
  Eq.(35)   nicely  holds   with  
$C_{\rm es}\cong 0.012 \sigma_1^3/\epsilon$ at $T=0.6$,  
while  a   closed  loop appears  at $T=0.2$. 
If  we set  $\epsilon/k_{\rm B}=100$ K and $\sigma_1=5~{\rm \AA}$, 
our $C_{\rm es}$ becomes  10 m$^4/$C$^2$. For 
ferroelectric polymers,    Eq.(35) 
was found with a negative coefficient \cite{Zhang,Z1,Furu1,Bl} 
($ -13.5$ m$^4/$C$^2$ after  electron irradiation\cite{Zhang}).
In contrast,  the piezoelectric relation 
($\bar{\gamma}\propto   P_z$) holds for relaxors 
 above the transition\cite{piezo}.

\section{ZFC$/$FC temperature changes}

A large number of ZFC$/$FC   experiments 
have been performed, where $T$ is varied 
at zero  or fixed ordering field (electric field\cite{West,Uesu,Z1,Ho1}, 
magnetic field\cite{F0,F1,F4}, and  stress\cite{F5,F3}).   
%For mixed crystals,   
%both of an  electric field\cite{Ho1}  and a strain \cite{F3} 
%can be  an ordering field.
However, the physical pictures of these processes remain  unclear.  
Here, we  show relevance  of 
collective, large-angle  orientational  changes in these cycles.

%15
\begin{figure}[tbp]
\begin{center}
\includegraphics[width=240pt]{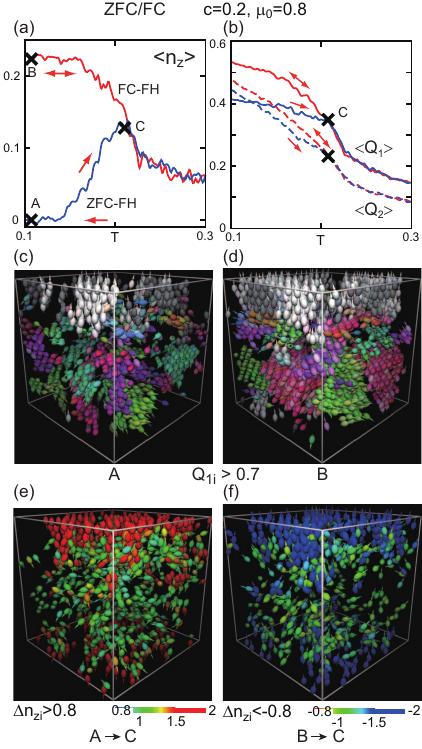}
\caption{ Results of  ZFC-FH and FC-FH thermal cycles 
for $c=0.2$ and  $\mu_0=0.8$, where 
 $\Delta\Phi=0.5$ on the paths of FC and FH 
and $T=0.1$ at points A and B. 
(a) $\av{n_z}=\av{\cos\theta}$ vs $T$. 
(b)  $\av{Q_1}$ and $\av{Q_2}$ vs $T$. 
Dipoles with $Q_{1i}>0.7$ 
are depicted at   A in (c) and at  B in (d). 
Also  depicted are  dipoles with 
large angle changes, where 
 $n_{zi}>0.8$ from A to C in (e) 
and $n_{zi}<-0.8$  from B to C in (f). 
Colors of these dipoles  are given according to the bars 
below these panels.  
}
\end{center}
\end{figure}

We followed cycles in Fig.15(a) 
setting $dT/dt=\pm 10^{-5}$ at fixed volume with $c=0.2$ and $\mu_0=0.8$. 
In ZFC-FH, (i) we cooled the system from a high-$T$  state 
to a low-$T$ state (point A)   with $\Delta\Phi={P}_z=0$ 
and then (ii) heated it with $\Delta\Phi=0.5$ ($E_{\rm a}=0.024$) 
back to the initial $T$. 
Subsequently, in FC-FH, (iii) we cooled the system to point B 
 with $\Delta\Phi=0.5$ 
and then (iv) heated it back with $\Delta\Phi=0.5$  fixed.
 We set $T=0.1$ at  A and B.

In (a), we  plot $\av{n_z}= {P}_z/\mu_0 n_1$ vs $T$ on the two paths.  
The two heating  curves 
meet at a freezing point C, where $T$ is given by $T_{\rm f}=0.21$ 
and the relaxation  time $\tau_1$ in Eq.(27) is of order $ 10^4$. 
This $T_{\rm f}$  is very close to $T_{\rm m}$ 
at $\omega=2\pi\times 10^{-4}$  in  Fig.6(a). 
Far below  $T_{\rm f}$, the two curves are 
largely separated  indicating marked  nonergodicity, 
while they coincide for $T>T_{\rm f}$ in the ergodic regime.   
%These curves closely 
%resemble the experimental ones in various systems. 

In (b), we display  
 $\av{Q_{\ell }}= \sum_{i\in 1}Q_{\ell i}/N_1$ 
($\ell=1,2$) in the same simulation run. From Eq.(24) they 
represent     the average  dipolar and  quadrupolar orders.  
The  difference  of $\av{Q_1}$   in the two cycles is 
at most $30\%$, while  that of  $\av{Q_2}$ is    only about    $5\%$. 
Note that   $Q_{2i}$ are rather insensitive to collective 
reorientations for most $i$ (see  Sec.VB).

In (c) and (d),  the dipoles with $Q_{1i}>0.7$ are depicted 
at   A  and  B.    These two patterns 
look similar, but  some PNRs  in the same locations  
in  A and B  have different polarization directions (for example, 
$[1\bar{1}\bar{1}]$ in A and $[\bar{1}11]$ in  B). 
In the present example, the potential 
energy $U$ is  $-5.86N\epsilon$ at A 
and  $-5.88N\epsilon$ at B. Their 
 difference ($= -0.02N\epsilon$) is small,   
but is still  5 times larger than 
  $-E_{\rm a} M_z (= -0.004 N\epsilon)$  
 at B (see  $U_{\rm d}$ in Eq.(11)).   
Note that  large potential barriers 
  exist  for reorientations of  PNRs  
from  the configurations at A to  those at B.  
These barriers  decrease with increasing $E_{\rm a}$, 
but its  present size 0.024 is small. 
If a  much larger $E_{\rm a}$  is applied 
at  A, there can be a transition to 
a ferroelectric state\cite{Bl,Bobnar}.

In (e) and (f), we  display  the dipoles with 
large angle  changes from A to C and from B to C.
They satisfy  
$\Delta n_{zi} (A\to C) = n_{zi}(C)-n_{zi}(A)>0.8 $  in (e)  and 
 $\Delta n_{zi} (B\to C) = n_{zi}(C)-n_{zi}(B)<-0.8$  in (f), 
where $n_{zi}$ are the $z$ component of ${\bi n}_i$.  
These large-angle changes are 
collective and heterogeneous. This  should be a universal feature 
 in  glass coupled with a phase transition.

On the two FH paths, the potential barriers 
between the two states   at the same $T$ remain  very large  
for   $T<T_{\rm f}$.  They can be overcome  by 
 thermal activations  at $T=T_{\rm f}$ (at C), where   the 
 reorientation rate  of  PNRs should be    comparable to 
 the inverse of the  observation time $\tau_{\rm obs}$. 
Estimating the former as the inverse of $\tau_1$ in Eq.(27) 
 and  setting $\tau_{\rm obs} \sim T/(dT/dt)$, we obtain 
\be 
 \tau_1 \sim  T/(dT/dt) \quad ({\rm at~point}~ C). 
\en 
Indeed, we have $\tau_1= 2.4\times 10^4 \cong  1.1 T/(dT/dt)$  at  C. 
Here,  $\tau_1$ at the freezing  should decrease 
significantly  for large  $E_{\rm a}$   (not shown here). 
It follows that  $T_{\rm f}$ at C decreases 
 with increasing  $\tau_{\rm obs}$. Note that   this dependence 
is   weak for   long  $\tau_{\rm obs}$ due to 
 the abrupt $T$ dependence of  $\tau_1$ at low $T$. 
It is well known that   nucleation in a metastable state 
starts  at an onset  temperature\cite{Onukibook}, which is rather 
well defined for long  $\tau_{\rm obs}$.

In Fig.15, we have used small $E_{\rm a}(= 0.024$).
However,  cooling at  a high electric field, 
Bobnar {\it et al.} \cite{Bobnar} 
detected  a field-induced ferroelectric 
transition.  Such  a phase  transition 
 can  be  predicted from a Ginzburg-Landau theory 
for $P_z$ including  the  field term $-E_{\rm a}P_z$  \cite{Bl,Onukibook}.

\section{ Antiferroelectric glass}

So far we have treated ferroelectric  glass.
However, antiferroelectric  
order has  been observed  
in  mixtures  containing  cyanide units CN$^-$   such as  KBr-KCN 
 at low $T$ 
\cite{Mertz,ori,Binder,anti-Loidl}. 
 It is also known that   antiparallel  alignment  freezes at low  $T$ 
 in polar  globular molecules such as cyanoadamantane \cite{anti,anti0}
containing CN 
%(C$_{10}$H$_{15}$CN) 
or betaine phosphate\cite{Albers} containing H$_3$PO$_4$
   due to their mutual steric hindrance. 
These systems    should become  antiferroelectric glass  at low $T$ 
 even without impurities. 
Here, we consider a mixture of 
dipoles and impurities introducing  a short-range 
interaction favoring antiparallel ordering. 

Supposing  top-tail  asymmetry of the dipoles, we  replace the factor $A_{ij}$ 
 ($i\in \alpha$ and $j\in \beta$) in Eq.(3) by 
\be 
{A}'_{ij}=A_{ij} +\delta_{\alpha 1}\delta_{\beta 1}J {\bi n}_i\cdot{\bi n}_j. 
\en 
The   second term yields an  exchange 
interaction  between dipoles $i$ and $j$,  
where  positive (negative) $J$ favors 
antiferroelectric (ferroelectric) ordering. 
We performed  simulation for $J=0.1$ with  
$c=0.2$, $\mu_0=0.8$,  and $\Delta\Phi=0$. 

In Fig.16(a), we plot $\av{Q_1}$ and $\av{Q_2}$. 
Here,  due to antiferroelectric ordering,  $\av{Q_1}$ 
remains very small at any $T$,  but  $\av{Q_2}$ increases up to 0.42 
with lowering $T$. 
Thus, the system exhibits quadrupolar order 
without dipolar order at zero  applied electric 
field \cite{Binder,ori}.  In  more detail, 
   we show   the  distribution functions 
$
P_\ell(Q_\ell)=\av{\sum_{i \in 1} \delta(Q_\ell-Q_{\ell i})/N_1} 
$  at $T=0.05$. 
%where we take the averages over all the dipoles 
%in a long time interval.  
In (b),   $P_1(Q_1)$ is nearly 
symmetric (even) with respect to $Q_1\to - Q_1$ 
and $P_2(Q_2)$ has a maximum at   $Q_2\cong 1$ 
leading to $\av{Q_2}\sim 0.4$.

Furthermore, in (c), 
a snapshot of the dipoles and the impurities is given, 
which looks very complicated. In (d), we show a typical 
antiferroelectric nanoregion  in the middle of the cell, which are  
viewed from two directions. 
Any dipole  $i$  in   this region 
satisfies $\bi n_i\cdot\bi n_j<-0.98$ 
for some nearby $j$ with $r_{ij}<1.4$ within the same region. 
It is composed of 170 
dipoles and surrounded by 130 impurities with 
 no impurities in its interior. 
In (d) and (e), cross-sectional  particle configurations 
are displayed at $z=H/2+2$ and $H/2+3$, respectively. 
We can see antiferroelectric ordering unambiguously  
for  the dipoles parallel or antiparallel 
 to the $z$ axis (perpendicular to 
$(111)$), while the orientations apparently look irregular 
for those perpendicular to $(\bar{1}11)$, 
$(1\bar{1}1)$, or $(11\bar{1})$.

%16 
\begin{figure}[tbp]
\begin{center}
\includegraphics[width=240pt]{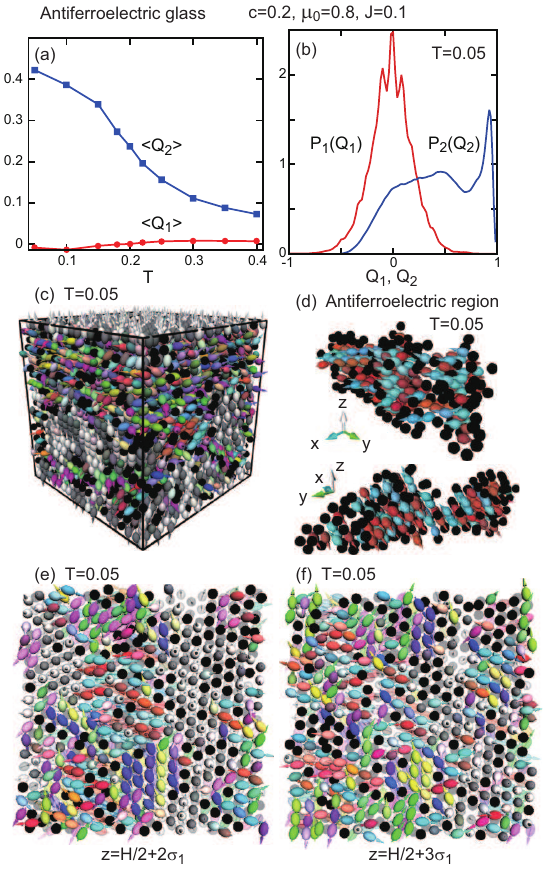}
\caption{
Antiferroelectric glass for $c=0.2$,   $\mu_0=0.8$, $\Delta\Phi=0$, 
and $J=0.1$, where $\av{Q_1}$ is small but  $\av{Q_2}$ increases 
gradually with lowering $T$ in (a).  
(b) Distributions  $P_\ell(Q_\ell)$ in Eq.(38)   at $T=0.05$, 
which  give $\av{Q_1}\cong 0$ and $\av{Q_2}\cong 0.42$.
 (c)  Snapshot of  dipoles   and impurities (black spheres) at $T=0.05$. 
(d) Antiferroelectric nanoregion  viewed from two directions. 
In (e) and (f), displayed are cross-sectional particle configurations 
at $z=H/2+2$ and $H/2+3$, respectively, at $T=0.05$. 
   }
\end{center}
\end{figure} 

\section{ Summary and remarks} 

With   molecular dynamics simulation, 
we have studied dipolar glass in mixtures of 
 dipolar spheroids  and apolar 
impurities in  applied electric field.
 Properly calculating the electrostatics, 
we have visualized polar nanoregions (PNRs)  
and clarified their role in the dielectric 
response.  We summarize 
our main results  as follows.

(i) In Sec.II, we have introduced 
  orientation-dependent  Lennard-Jones potentials 
 mimicking   spheroidal repulsion. For  its mild aspect ratio, 
the particles first form 
a fcc plastic  crystal. Then, at lower $T$,  
 the spheroids   align along  
$\av{111}$   resulting in  rhombohedral structures. 
Assuming that each  spheroid has a dipole 
 parallel to its long  axis, 
we have constructed an 
 electrostatic energy $U_{\rm d}$ in Eq.(11), which accounts  for 
the image dipoles and the  applied field 
$E_{\rm a}$. In equilibrium, the differential 
 susceptibility $\chi_{\rm dif}=dP_z/dE_{\rm a}$ 
is related to the 
polarization fluctuations as in Eq.(17).

(ii) In Sec.III, we have  presented  results on a  
 structural phase transition in a one-component 
system of  dipolar spheroids.  
It changes from a fcc crystal to a  polycrystal 
with eight rhombohedral variants. 
This transition  occurs  in a narrow temperature range 
due to the finite size effect imposed by the  metal walls.

(iii)  In Sec.IV, we have examined   diffuse ferroelectric 
 transitions.  The impurity distribution has been determined during 
crystallization, so marked  impurity 
clustering has appeared.  
In our model, ferroelectric domains 
are broken up into smaller PNRs with increasing the impurity 
concentration $c$.  
For $c=0.2$, we have calculated the 
orientational time correlation function $C_1(t)$ in Fig.5 
and the   dielectric permittivity  
in Fig.6. 
The temperature of maximum of $\ve'$ is  written as $T_{\rm m}(\omega)$. 
For very small $\omega$, the polarization fluctuations are  enhanced  
for  $T>T_{\rm m}$, but  are composed of frozen PNRs  and 
thermal fluctuations for  $T<T_{\rm m}$. 
 Individual PNRs have been visualized in Fig.10. 
 The surface effects on the dipole orientations  
and the local electric fields    have   been 
examined in Sec.IVF and Appendix A. 

(iv) In Sec.V, we have examined  the 
polarization and the strain to cyclic applied electric field. 
%The results  resemble  those in the previous experiments. 
At relatively high $T$, there is no hysteresis and an 
electrostriction relation holds. 
At low $T$, the polarization is on a hysteresis loop. 
In the cycle,   collective  large-angle changes are dominant 
where    $\chi_{\rm dif}= dP_z/dE_{\rm a}$ is large.  

(v) In Sec.VI, we have investigated the ZFC-FH  and FC-FH  thermal 
cycles  in accord  with  the 
previous experiments. The  frozen states 
at the lowest $T$ in the two cycles 
 have been  visualized in Fig.15. 
On the FH paths, heterogeneous collective reorientations 
have been found. These paths meet 
at a  temperature $T_{\rm f}$, 
at which the reorientation rate ($\sim \tau_1^{-1})$ 
is of the same order as the ramping rate of the temperature 
($\sim (dT/dt)/T$). 

(vi) In Sec.VII, we have investigated 
 antiferroelectric glass  
by introducing a short-ranged exchange interaction 
stemming from molecular shape asymmetry. 
We have visualized a typical  antiferroelectric nanoregion.\\
(vii) In Appendix B, we have 
shown the method of calculating $\ve'$ and $\ve''$ and 
  found their algebraic behavior 
$(\propto \omega^{-\beta})$  
at relatively large $\omega$ in the ergodic $T$ range.

Finally, we   remark  on future problems. 
(1) The isochoric  specific heat $C_V$ can be calculated from 
the  average energy. We found that it has  a rounded peak in our 
mixture systems (not shown in this paper). 
This is  consistent with the behavior of the 
 isobaric specific heat $C_p$ in previous experiments 
\cite{ori,Mertz,Kawaji,Tachibana}.
(2) There is a gradual  crossover 
in the polarization fluctuations 
 in  the diffuse  transition. 
For example, the PNRs have no 
clear boundaries at relatively high $T$,  while  
sharp interfaces can appear at low  $T$. 
It is of interest how  the space correlations  in the polarization and  the  
 particle displacements depend on $T$.  
(3)  In real systems,  impurities or mixed components 
have charges or dipoles. 
In solids,  the polarization response can be  large 
 when  ion displacements  occur   within unit cells 
  as a phase transition. 
These features  
should be accounted for in  future simulations. 
(4) 
Intriguing critical dynamics exists  
in the ergodic $T$ range\cite{Klee1,Bokov3,Bokov4,Cowley}, 
as suggested by Eq.(31). 
The aging and memory effects at low $T$\cite{aging0,aging} 
should also be studied in future
(see the last paragraph of Sec.VB).

\noindent 
{\bf Acknowledgments}\\
This work was supported by 
%KAKENHI No. 25610122, 
  KAKENHI 15K05256, and KAKENHI 25000002.
The numerical calculations were preformed on
CRAY XC40 at YITP in Kyoto University
and on SGI ICE XA/UV at ISSP in the University of Tokyo.

%\vspace{2mm}
\noindent{\bf Appendix A: Electrostatics of dipole systems}\\
\setcounter{equation}{0}
\renewcommand{\theequation}{A\arabic{equation}}

Here, we explain the electrostatics  of  dipoles 
between metal walls in applied field 
\cite{Takae1,Takae2,Hautman,apply,Klapp}.  
The electric potential due to 
the image dipoles is equivalent to that due to  the surface 
charge densities, written as 
$\sigma_0(x,y)$ at $z=0$  and $\sigma_H(x,y)$ at $z=H$. 
Without adsorption and ionization on the surfaces, 
the dipole centers are 
somewhat away from the walls (see the comment below  Eq.(6)). Then,    
\be 
4\pi \sigma_0=  {E_z(x, y, 0)},\quad 
4\pi\sigma_H= - {E_z(x, y,H)},
\en  
where   $E_z =-\p \Phi/\p z$. 
We consider the  2D Fourier expansions of $\sigma_\lambda$. 
For $\lambda=0$ and $H$ they are 
 \be 
 \sigma_\lambda({\bi r}_\perp)={\bar \sigma}_\lambda + 
\sum_{{\bi k}\neq{\bi 0}}\sigma_{\lambda {\bi k}}
\exp[{{\rm i}{\bi k}\cdot{\bi r}_\perp}],
\en  
where    ${\bi r}_\perp=(x,y)$,  and 
${\bi k}=(2\pi /L)(n_x,n_y)\neq (0,0)$ with $n_x$ 
and $n_y$ being integers.  The first term 
is the  mean surface charge density 
${\bar\sigma}_\lambda= \int_0^L  dx\int_0^L 
 dy~ \sigma_\lambda (x,y) /L^2$. From Eq.(10) we can 
express the Fourier components $\sigma_{\lambda{\bi k}}$ as\cite{apply}   
\be 
\sigma_{\lambda{\bi k}}=-  
\sum_{j}({\bi\mu}_j\cdot\nabla_j)
[G_k^\lambda(z_j) 
e^{-{\rm i}{\bi k}\cdot{\bi r}_j}]/L^2, 
\en 
where $\nabla_j= \p/\p{\bi r}_j$, 
$G_k^0(z)= {\sinh(k(H-z))}/\sinh(kH)$,  and $G_k^H(z)= {\sinh(kz)}/\sinh(kH)$ 
with  $k= |{\bi k}|$.

%17
\begin{figure}
\begin{center}
\includegraphics[width=250pt]{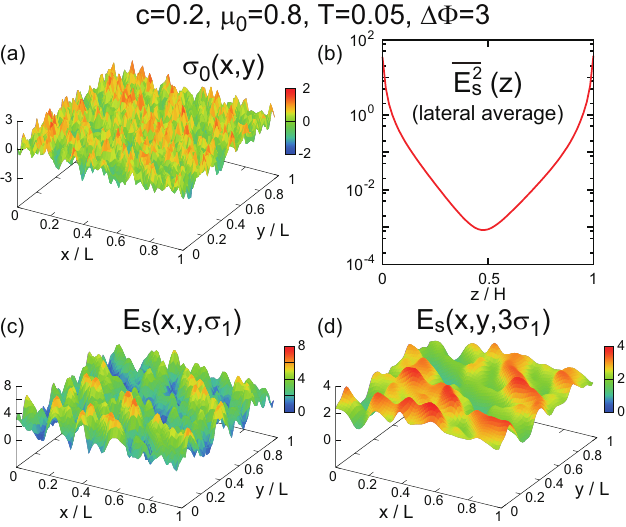}
\caption{ Surface charge effects in  
ferroelectric glass  for $\mu_0=0.8$, $c=0.2$,  $T=0.05$, 
and  $\Delta\Phi=3$. (a)  $\sigma_0(x,y)$ on the $xy$ plane at $z=0$ 
exhibiting both microscopic and mesoscopic fluctuations, 
(b) Lateral average $\overline{E_{\rm s}^2}$  in Eq.(A12) vs $z/H$.  
(c) $E_{\rm s}$ on the $xy$ plane at $z=1$,  and 
(d) that at $z=3$.  Here, 
$E_{\rm s}= |\nabla\phi_{\rm s}|$ arises from the surface charge deviations 
 decaying far from the walls, 
but the nanodomain contribution to 
$E_{\rm s}$ exceeds  the microscopic part   with increasing $z$. 
}
\end{center}
\end{figure}

For dipolar systems, the  Poisson equation is written as 
\be 
\nabla^2 \Phi= 4\pi \nabla\cdot 
\sum_i {\bi \mu}_i\delta({\bi r}-{\bi r}_i). 
\en   
 Integration of Eq.(A4) in the cell 
 yields ${\bar{\sigma}}_0+{\bar{\sigma}}_H=0$. 
We also multiply   Eq.(A4) by $z$  and integrate it  
in the cell. Using the total polarization $M_z$ we 
 find\cite{Hautman,Takae1}
\be 
\bar{\sigma}_0= -\bar{\sigma}_H 
=  E_{\rm a}/4\pi+M_z/V ,
\en  
without surface adsorption and ionization. 
The fluctuations of ${\bar{\sigma}}_0$ and $M_z/V$  
thus coincide  at fixed $E_{\rm a}$ \cite{Hautman,apply,Takae1,Takae2}.

The mean surface charge densities 
produce the potential $-4\pi {\bar\sigma}_0z$ in the cell, so  
 $\Phi$ consists of three parts as  
\be 
\Phi({\bi r}) = \Phi_{\rm d} ({\bi r})  -4\pi {\bar\sigma}_0z+ 
\phi_{\rm s}({\bi r}), 
\en 
which is equivalent to Eq.(10). 
The first term   $\Phi_{\rm d}$ arises from the dipoles in the cell. 
Imposing  the lateral periodic boundary condition, we express it as   
\be 
\Phi_{\rm d}  ({\bi r}) = \sum_{{\bi m}_\perp} 
{\sum_i}  {\bi g}( {{\bi r}-{\bi r}_{i} + 
L{\bi m}_\perp})\cdot{\bi \mu}_i,
\en   
where  ${\bi g}({\bi r})=r^{-3}{\bi r}$ 
and $  {\bi m}_\perp= (m_x,m_y,0)$ 
with $m_x$ and $m_y$ being integers.
The third term $\phi_{\rm s}$ in Eq.(A6)  
arises from the charge density deviations  
$\delta\sigma_\lambda(x,y)=\sigma_\lambda -{\bar{\sigma}}_\lambda$.  
In terms of $\sigma_{\lambda{\bi k}}$ in Eq.(A2), 
$\phi_{\rm s}$  is expressed as 
\be 
\phi_{\rm s}=\frac{2\pi }{L^2}
\sum_{{\bi k}\neq {\bi 0}}
\frac{1 }{k}e^{{\rm i}{\bi k}\cdot{\bi r}_\perp}
\bigg[\sigma_{0{\bi k}}e^{-kz}+\sigma_{H{\bi k}}e^{-k(H-z)}
\bigg]. 
\en
Now  the local electric field ${\bi E}_i$ is written as 
\be 
{\bi E}_i= {\bi E}_i^{\rm d} + {\bi E}_i^{\rm sur}.
\en 
The first term  arises  from the other dipoles in the cell: 
\be 
{\bi E}_i^{\rm d}=  -\sum_{{\bi m}_\perp} {\sum_{j}}' 
{\ten{\cal T}} ({{\bi r}_{ij} +  L{\bi m}_\perp})  \cdot  
{{\bi \mu}_j },
\en 
The second term is due to  the surface charges: 
\be 
{\bi E}_i^{\rm sur}= 
-4\pi {\bar\sigma}_0{\bi e}_z + 
{\bi E}_{\rm s}({\bi r}_i),
\en 
where   the first term is homogeneous 
and ${\bi E}_{\rm s}({\bi r})= -\nabla\phi_{\rm s}({\bi r})$ is due to 
$\delta{\sigma_\lambda}= {\sigma_\lambda}-{{\bar\sigma}_\lambda}$. 
The dipoles next to the walls  are  parallel or antiparallel 
to the $z$ axis due to ${\bi E}_{\rm s}$ even for  $\Delta\Phi=0$ 
(see  the  snapshots   in this paper)\cite{apply}.  
 However, as in Fig.17(b), ${\bi E}_{\rm s}$  is  negligibly small 
(even in ferroelectric states) 
if the distances from the walls  exceed the typical domain size. 
This  is due to the factors $\exp({-kz})$ and $\exp(-k(H-z))$ 
in Eq.(A8).

In Fig.17,   we show   (a)$\sigma_0(x,y)$, 
 (b)$\overline{E_{\rm s}^2} (z)$, (c)$E_{\rm s}(x,y,1)$, and  
(d)$E_{\rm s}(x,y,3)$  in ferroelectric glass of our system, where we set 
$E_{\rm s}(x,y,z)=|{\bi E}_{\rm s}|= |\nabla\phi_{\rm s}|$ and  
\be 
\overline{E_{\rm s}^2} =\int_{0<x,y<L} dx dy E_{\rm s}(x,y,z)^2/L^2.
\en 
Here,   $\sigma_0(x,y)$ in (a) and  $E_{\rm s}(x,y,1)$ in (c)  
consist of microscopic and mesoscopic fluctuations. 
  The latter arise from the PNRs  near the surface 
 from Fig.11, being apparent  in (d). 
For $z$ longer than the  PNR length, 
$E_{\rm s}(x,y,z)$  decays to zero in (b). 
Thus,  
 ${\bi E}_i^{\rm sur} \to -4\pi {\bar\sigma}_0{\bi e}_z$ 
far from the walls, which was previously 
found  for liquid water\cite{Takae1,Takae2}.

\vspace{2mm}
\noindent{\bf Appendix B: Linear response to 
oscillating   field and frequency-dependent susceptibilities
}\\
\setcounter{equation}{0}
\renewcommand{\theequation}{B\arabic{equation}}

%18  
\begin{figure}[t]
\begin{center}
\includegraphics[width=230pt]{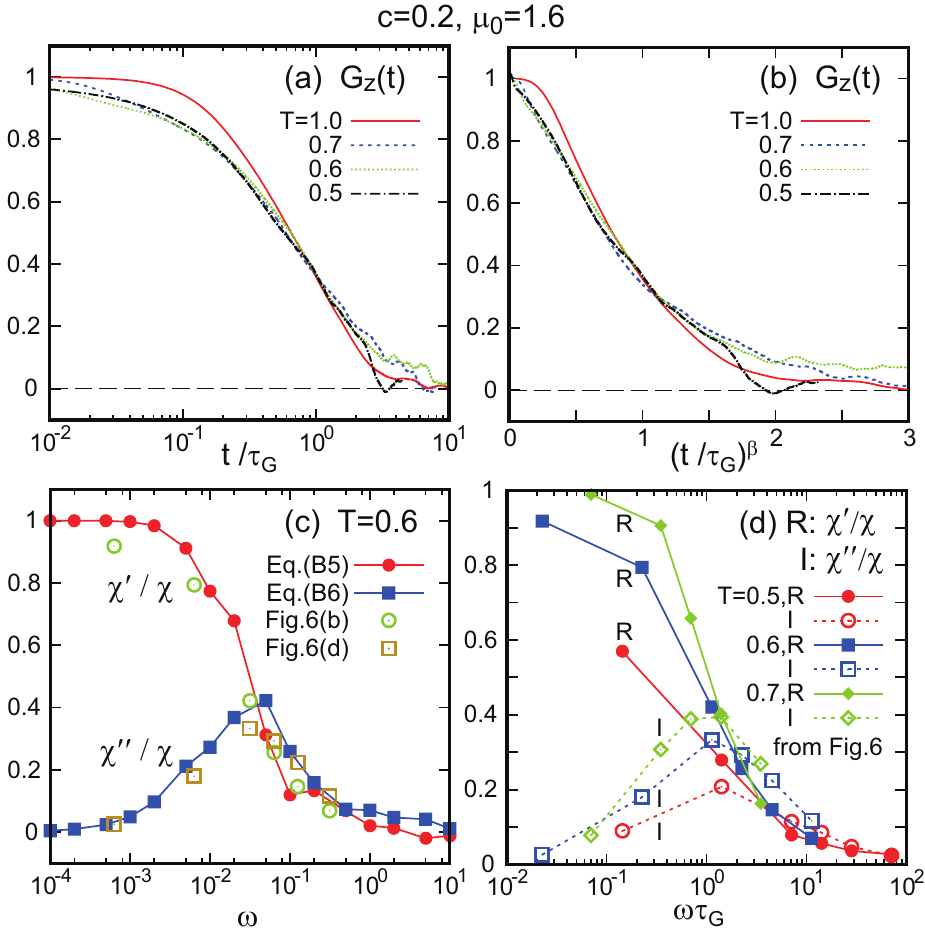}
\caption{ 
Dielectric relaxation for $c=0.2$ and $\mu_0=1.6$.  
(a)  $G_z(t)$ in  Eq.(B4) vs $t/\tau_G$ at $T=0.5,0.6, 0.7$, and 1.0 on 
 a semi-logarithmic scale, 
where $\tau_G$ is determined by $G_z(\tau_G)=e^{-1}$. 
(b) $G_z(t)$  vs $(t/\tau_G)^\beta$ with $\beta=0.57$ on
a linear scale. (c)  $\chi'(\omega)/\chi$  and 
$\chi''(\omega)/\chi$ vs $\omega$ from  one-sided Fourier transformation 
of $G_z(t)$ (filled symbols) and those from  data in 
Fig.6 (open symbols). (d) 
  $\chi'(\omega)/\chi$ (R)  and $\chi''(\omega)/\chi$ (I) vs $\omega\tau_G$ 
at $T=0.5$, 0.6, and 0.7 from  data in  Fig.6.
}
\end{center}
\end{figure}

We  applied  a  small sinusoidal electric field  
of the form $ E_{\rm a}(t)= E_{\rm a}^1 \sin (\omega t)$ 
with $ E_{\rm a}^1= 0.047$. 
We  calculated the polarization response $\delta P_z=
\delta M_z/V$ to this perturbation over 10 periods. 
After  a few periods, it is   expressed as   
\be 
\delta{P_z}(t)  =[\chi'(\omega)\sin (\omega t) -
\chi''(\omega)\cos(\omega t)] E_{\rm a}^1 ,
\en 
where $\chi'$ and $\chi''$ are the frequency-dependent 
susceptibilities.  Then, $\ve'$ and $\ve''$  in Fig.6 are defined by  
\be 
\ve'= 1+4\pi \chi', \quad \ve''=4\pi \chi''.
\en 
The Hamiltonian ${\cal H}$ increases  as 
$ {\overline{d{ \cal H}/dt}}
= V\omega\chi''|E_{\rm a}^1|^2/2$ in time (see below Eq.(20)), 
where the time average is taken in one period.  
From  Eq.(A5)   the mean  surface charge density at the bottom wall is 
written as  
\be 
\delta \av{{\bar\sigma_0}}(t) = [\ve' 
\sin (\omega t) - \ve'' \cos(\omega t)] E_{\rm a}^1 /4\pi  , 
\en 
which oscillates 
 as $\sin (\omega t-\delta_{\rm p})$ with $\tan\delta_{\rm p}= \ve''/\ve'$. 
 In  Fig.6,  we give  the resultant  $\ve'$ and $\ve''/\ve'$ 
in a wide $T$ range including the nonergodic range.

On the other hand,  around  equilibrium, 
we can use the linear response theory \cite{Kubo} 
 for the Hamiltonian 
of the form (15). Within this scheme, the dielectric response can be expressed 
in terms of the time-correlation function for the deviation  
$\delta M_z(t)=M_z(t)-{\av{M_z}}_{\rm e}$: 
\be 
G_z(t)= {\av{\delta M_z(t+t_0)\delta M_z(t_0)}}_{\rm e}/Vk_{\rm B}T\chi ,
\en 
where ${\av{\cdots}}_{\rm e}$ represents  
the equilibrium average and 
$\chi= {\av{(\delta M_z)^2}}_{\rm e}/Vk_{\rm B}T=
(\ve-1)/4\pi$ at $E_{\rm a}=0$ (see Eq.(17)). 
Using  $G_z(t)$ we obtain the linear response relations, 
\bea
\chi'(\omega)/\chi &=& 1-  \omega\int_0^\infty dt G_z (t) \sin(\omega t),\\
 \chi''(\omega)/\chi &=&  \omega\int_0^\infty dt G_z (t) \cos(\omega t).  
\ena 
The  complex susceptibility  $\chi'-i\chi''$  can be expressed as 
 \be 
\chi'(\omega)-i\chi''(\omega)= -\chi \int_0^\infty dt {\dot G}_z(t) 
e^{-i\omega t}, 
\en 
in terms of   the time derivative ${\dot G}_z(t)= {dG_z(t)}/{dt }$.

In  Fig.18(a), we show our numerical 
results of $G_z(t)$  in Eq.(B4) at $T=0.5, 0.6,0.7$, and 1.0 
for $c=0.2$, $\Delta\Phi=0$, and $\mu_0=1.6$, where 
the data at long times are inaccurate, however. 
From  $G_z(\tau_G)=e^{-1}$, we 
 define  the relaxation time 
 $\tau_G$, which   is somewhat shorter than  $\tau_1$ in Fig.5. 
In fact, we  obtain $(\tau_G, \tau_1)= (230, 1300), (36,60)$, $(11,17)$, 
and $(1.8, 2.5)$ for $T=0.5, 0.6$, 0.7, and 1.0, respectively. 
We  may introduce  another   time by  
$\tau'_G=\lim_{\omega\to 0}\chi''/\omega\chi=\int_0^\infty dt G_z(t)$, but 
we confirm $\tau_G'\sim  \tau_G$.

In (b), the initial decay of $G_z(t)$ is  well  fitted to  
\be 
G_z(t)=1- A_{\rm p} (t/\tau_G)^\beta +\cdots \quad (t\ls \tau_G), 
\en 
where $\beta\cong 0.57$ and $A_{\rm p}\cong 0.63$.  Then,  ${\dot G}_z(t) 
\propto -t^{\beta-1} $ for $t\ls \tau_G$. If this 
 is substituted  into  Eq.(B7), we find   
\be 
{\chi'(\omega) }
- i\chi''(\omega) \cong  B_{\rm p}  e^{-i \pi \beta/2}
(\omega\tau_G)^{-\beta} \quad (\omega\tau_G\gs 1),
\en  
where  $B_{\rm p}= \beta \Gamma(\beta)A_{\rm p}\chi\sim \chi$. 
The  algebraic form (B9) with $0<\beta<1$   has  been 
observed in many systems 
including relaxors and mixed crystals \cite{Bokov3,Klee,ori}.

We calculated the ratios   $\chi'(\omega)/\chi$  and 
$\chi''(\omega)/\chi$ from Eqs.(B5) and (B6)
using  $G_z(t)$ in (a). 
In (c), they are  plotted at $T=0.6$  
 together with   those  from the data   in Fig.6, 
where the latter are  from Eq.(B1). 
 The  points  from these  two  sets fairly 
agree  for any $\omega$. 
In (d), we also plot  these ratios   
vs $\omega\tau_G$  at three temperatures using the results  in  Fig.6.
The behaviors in the region $\omega \tau_G \gs 1$ 
in (c) and (d)  support  the algebraic   form  (B9). 
%though the calculations  in these two methods are still crude.

\end{document}